\documentclass[
aip,
dcolumn,
bm,
nofootinbib,
notitlepage]{revtex4-1}
\usepackage{graphicx}
\usepackage{color}
\usepackage[latin1]{inputenc}
\usepackage[T1]{fontenc}
\usepackage{amsmath}
\usepackage{amssymb}
\usepackage{braket}
\usepackage{mathrsfs}
\usepackage{hyperref}
\usepackage{upgreek}
\usepackage{pstricks-add}
\usepackage{natbib} 
\usepackage{listings}
\usepackage{tikz}
\usetikzlibrary{mindmap}

\begin{document}
\title{Cavity-Altered Thermal Isomerization Rates and Dynamical Resonant Localization in Vibro-Polaritonic Chemistry}
\author{Eric W. Fischer}
\affiliation{Theoretische Chemie, Institut f\"ur Chemie, Universit\"at Potsdam,
Karl-Liebknecht-Strasse 24-25, D-14476 Potsdam-Golm, Germany}

\author{Janet Anders}
\affiliation{Institut f\"ur Physik und Astronomie, Universit\"at Potsdam,
Karl-Liebknecht-Strasse 24-25, D-14476 Potsdam, Germany and CEMPS,
Physics and Astronomy, University of Exeter, Exeter EX4 4QL,
United Kingdom}

\author{Peter Saalfrank}
\affiliation{Theoretische Chemie, Institut f\"ur Chemie, Universit\"at Potsdam,
Karl-Liebknecht-Strasse 24-25, D-14476 Potsdam-Golm, Germany}

\let\newpage\relax

\begin{abstract}
It has been experimentally demonstrated that reaction rates for molecules embedded in microfluidic optical cavities are altered when compared to rates observed under ``ordinary'' reaction conditions. However, precise mechanisms of how strong coupling of an optical cavity mode to molecular vibrations affect the reactivity and how resonance behavior emerges are still under dispute. In the present work, we approach these mechanistic issues from the perspective of a thermal model reaction, the inversion of ammonia along the umbrella mode, in presence of a single cavity mode of varying frequency and coupling strength. A topological analysis of the related cavity Born-Oppenheimer potential energy surface in combination with quantum mechanical and transition state theory rate calculations reveals two quantum effects, leading to decelerated reaction rates in qualitative agreement with experiments: The stiffening of quantized modes perpendicular to the reaction path at the transition state, which reduces the number of thermally accessible reaction channels, and the broadening of the barrier region which attenuates tunneling. We find these two effects to be very robust in a fluctuating environment, causing statistical variations of potential parameters such as the barrier height. Further, by solving the time-dependent Schr\"odinger equation in the vibrational strong coupling regime, we identify a resonance behavior, in qualitative agreement with experimental and earlier theoretical work. The latter manifests as reduced reaction probability, when the cavity frequency $\omega_c$ is tuned resonant to a molecular reactant frequency. We find this effect to be based on the dynamical localization of the vibro-polaritonic wavepacket in the reactant well. 
\end{abstract}

\let\newpage\relax
\maketitle

\section{Introduction}
Recent experiments have shown that when molecules are placed in 
reflecting nanoscale cavities such as Fabry-P{\'e}rot cavities,
the quantized electromagnetic cavity modes and their possible strong coupling to molecular electronic or vibrational excitations, can affect 
the chemical reactivity, spectroscopic response, and other molecular properties\cite{george2015,ebbesen2016,kockum2019,herrera2020}.
\vspace{0.2cm}
\\
Focusing on Vibrational Strong Coupling (VSC) and its effect on reaction kinetics here, we motivate our work by referring to celebrated experiments of Ebbesen and coworkers\cite{thomas2019}. These authors demonstrated for silane molecules with cleavable Si-C and Si-O bonds, that for most temperatures in a range between 25 and 30$^o$ C and for most cavity mode frequencies studied, rates for thermal breaking of either Si-C or Si-O were significantly decelerated when molecules were inside a microfluidic Fabry-P{\'e}rot cavity, compared to the outside-cavity scenario. 
Second, by tuning the frequency of the cavity mode around the Si-C or Si-O stretch vibrational frequencies, a resonance behavior was observed, {\em i.e.}, the total reaction rate (cleavage of the molecule) was strongly decelerated. 
Third, through tuning the cavity frequency, it was possible to enforce bond-selective chemistry, {\em i.e.}, preferential Si-C or Si-O breaking, by suppressing the concurring reaction. Using Arrhenius plots $\ln\,k$ {\em vs.} $1/T$, where $k$ is the rate for either Si-C or Si-O bond breaking and $T$ the temperature, it was also possible to deduce activation free energies, $\Delta G^\ddagger$, from Eyring's transition state theory.
It was found that, $\Delta G^\ddagger$ is strongly increased in the cavity, by an amount depending on cavity mode frequency and reaction (Si-C or Si-O bond cleavage), leading to reduced rates. When na{\"i}vely correlating activation free energies with energy differences between transition states and reactants, one may speculate that 
the reduced reactivity is due to cavity-enhanced classical activation energies. This is not the case as we shall argue later.
\vspace{0.2cm}
\\
Several theoretical efforts have been made to explain the reactivity of molecules in cavities\cite{galego2015,flick2017a,flick2017b,martinez2018,lacombe2019,angulo2019,galego2019,schaefer2019,triana2020,angulo2020,li2020a,li2020b,zhdanov2020,li2021b,li2021c,li2021d,du2021,lihuo2021a,fischer2021} (and related effects) under VSC. A cavity Born-Oppenheimer (cBO) approximation has been formulated in which the molecular electronic degrees of freedom separate from the nuclear and cavity degrees of freedom, and the concept of potential energy surface (PES) was extended to molecular cavity quantum electrodynamics (QED) \cite{galego2015,flick2017a,flick2017b}. This allows to combine quantum chemistry to calculate potential energy surfaces and dipole functions with cavity QED. In the cBO approximation, the cavity PESs (cPESs) depend on both the relevant molecular degrees of freedom and on the cavity modes. 
\vspace{0.2cm}
\\
Ground-state cPES have been computed in Refs.\cite{flick2017b,galego2019,triana2020}, for example. It was argued that equilibrium bond lengths are influenced by VSC due to cavity-distorted PESs\cite{galego2015,flick2017b,triana2020}, or that energy barriers and, as a consequence, reaction kinetics are affected \cite{galego2019}. This is in qualitative agreement with the experiments reported in Ref.\cite{ebbesen2016,thomas2019}, observing cavity-dependent activation free energies and rates. We already note here that in some of these works\cite{galego2015,galego2019,triana2020}, Hamiltonians were used, that lack the dipole self-energy (DSE) term in the vibration-cavity interaction. This leads to large differences already for classical activation energies inside and outside the cavity in contrast to the situation where the DSE is included \cite{lihuo2021a,fischer2021}.
\vspace{0.2cm}
\\
Moreover, in a very recent work, Sch\"afer \textit{et al.}\cite{schaefer2021} theoretically predicted a dynamical resonance effect for a S$_N$2-reaction, leading to 
a reduced reaction rate near a frequency resonant with a molecular vibration, in accordance with experiment\cite{thomas2016}. The authors of Ref.\cite{schaefer2021} simulated the reaction with a ground-state multi-mode model using a quantum-classical approach, where the time evolution of the nuclei was treated by Ehrenfest dynamics. It was argued that the resonance effect emerges due to a cavity-mediated collective energy redistribution in the molecular vibrational degrees of freedom, which effectively lowers the energy in the reactive mode and therefore suppresses the reaction.
\vspace{0.2cm}
\\
Finally, in Ref.\cite{lihuo2021a} the hydrogen-exchange between an H-donor and  H-acceptor molecule was treated in a one-dimensional, symmetric double-well potential model, resulting in a two-dimensional cPES when strongly coupled to an optical cavity mode. This model, which is in close analogy to our own previous treatment of the ammonia inversion reaction in Ref.\cite{fischer2021}, showed also a dynamical resonant suppression of transfer rates. In this case the resonance had a different character, with cavity frequencies close to the (imaginary) frequency at the transition state being efficient in suppressing the reaction. In Ref.\cite{lihuo2021a}, the kinetics was treated beyond transition state theory, but still classically using Grote-Hynes rate theory. 
\vspace{0.2cm}
\\
In this work, we consider in detail the effect of strong coupling of a cavity mode to a molecular reaction coordinate on reaction kinetics. As in Ref.\cite{fischer2021}, a one-dimensional ammonia inversion model in a symmetric double-well, coupled to a single cavity mode, is adopted as a simple model system. In contrast to most previous works (see Ref.\cite{galego2019} for an exception), full quantum rate theory, which accounts for both tunneling and the effects of quantization (of cavity and molecular modes), is adopted besides Eyring transition state theory. 
Using the so-called Pauli-Fierz Hamiltonian \cite{flick2017a,flick2017b,schaefer2018} including the dipole self-energy as in Refs.\cite{lihuo2021a,fischer2021}, we find reduced inversion rates under VSC conditions. We argue that the observed reduction of reaction rates is dominated by quantum effects (effects of vibrational
quantization and suppressed tunneling). Within rate theories based purely on classical  activation energies, no effect on rates would be obtained -- in agreement with 
findings in Refs.\cite{lihuo2021a,fischer2021}. 
Further, using a time-dependent description, we find a dynamical resonance effect at 
cavity frequencies close to a vibrational transition frequency of the reactant,
with reactants localizing and the reaction being suppressed. In extension to recently proposed\cite{schaefer2021} cavity-enhanced intramolecular vibrational energy redistribution (IVR) as the main trapping mechanism at resonance, we argue that the cavity mode itself can act as an efficient energy acceptor leading to reduced reaction  rates.
\vspace{0.2cm}
\\
The paper is organized as follows. In Sec.\ref{theory}, we recall our treatment of anharmonic vibro-polaritons within the Pauli-Fierz formalism \cite{fischer2021} (Sec.\ref{theorya}) and of the  model reaction and parameters (ammonia inversion, Sec.\ref{model}). In Sec.\ref{subsec.theory_thermal_rates}, we describe the quantum and transition state rate theories employed here. In Sec.\ref{results}, we present and discuss results for the ammonia inversion: general topological features of the cPES for various coupling strengths and cavity frequencies (Sec.\ref{subsec.cpes_topology}) and corresponding reaction rates, calculated from stationary rate theory (Sec.\ref{subsec.results_thermal_rates}). In Sec.\ref{subsec.resonant_dynamical_localization}, a resonance phenomenon is discussed in the context of ``dynamical localization'' along the molecular coordinate, using the time-dependent Schr\"odinger equation. Sec.\ref{sec.conclusion} summarizes this work and addresses aspects to be covered in the future.
\section[h!]{Methods and Model}
\label{theory}
\subsection{Anharmonic Vibrational Polaritons: Pauli-Fierz Hamiltonian and cPESs}
\label{theorya}
We consider a molecule  with a single (reaction) mode coupled to a single cavity mode. The corresponding vibrational Pauli-Fierz Hamiltonian in length gauge and dipole approximation is given by
\begin{equation}
\hat{H}
=
\hat{H}_\mathrm{S}
+
\hat{H}_\mathrm{C}
+
\hat{H}_\mathrm{SC} \quad ,
\label{eq.vib_pauli_fierz_hamilton}
\end{equation}
based on a cavity Born-Oppenheimer (cBO) type perspective for the electronic ground state\cite{flick2017a,flick2017b,schaefer2018,lihuo2021a,fischer2021}. Here, $\hat{H}_\mathrm{S}$ is the vibrational system Hamiltonian given by
\begin{equation}
\hat{H}_\mathrm{S}
=
-
\dfrac{\hbar^2}{2\mu}
\dfrac{\partial^2}{\partial q^2}
+
V(q)
\label{eq.molecular_system}
\end{equation}
with reduced mass $\mu$, molecular coordinate $q$ and an anharmonic, molecular ground state Born-Oppenheimer potential energy surface $V(q)$. Further, $\hat{H}_\mathrm{C}$ is the harmonic, single-cavity mode Hamiltonian, which reads in coordinate representation
\begin{equation}
\hat{H}_\mathrm{C}
=
\dfrac{1}{2}
\left(
\hat{p}^2_c
+
\omega^2_c\,x^2_c
\right) \quad ,
\end{equation}
where $\omega_c$ is the frequency of the cavity mode, and $x_c=\sqrt{\frac{\hbar}{2\omega_c}}(\hat{a}^\dagger+\hat{a})$ (in atomic units of $\sqrt{m_e}\,a_0$) and $\hat{p}_c={i}\,\sqrt{\frac{\hbar\omega_c}{2}}(\hat{a}^\dagger-\hat{a})$ are cavity mode (``photon'') coordinate and momentum operators, respectively. $\hat{a}^\dagger/\hat{a}$ are the bosonic photon creation/annihilation operators satisfying $[\hat{a},\hat{a}^\dagger]=1$. 
\vspace{0.2cm}
\\
The third term in Eq.\eqref{eq.vib_pauli_fierz_hamilton}, $\hat{H}_\mathrm{SC}$, is given for a cavity mode polarized along the system coordinate $q$ by\cite{fischer2021}
\begin{equation}
\hat{H}_\mathrm{SC}
=
\sqrt{\dfrac{2\omega_c}{\hbar}}\,g\,x_c\,d(q)
+
\dfrac{g^2}{\hbar\omega_c}
d^2(q) \quad .
\label{hsc}
\end{equation}
Here, $g$ is a coupling parameter with dimension of an electric field strength and $d(q)$ is the nonlinear molecular dipole function along the reaction coordinate, $q$.  Further, $g$ can be directly connected to a dimensionless coupling parameter $\eta$, or alternatively to a Rabi frequency $\Omega_R$, as\cite{kockum2019,fischer2021}
\begin{equation}
g
=
\dfrac{\hbar\omega_c}{\vert d_{fi}\vert}\,\eta  = \dfrac{\hbar\Omega_R}{2 |d_{fi}|}
 \quad ,
\label{eq.g_vs_eta_relation}
\end{equation}
with vibrational transition dipole matrix element $d_{fi}$ connecting two molecular vibrational eigenstates, $i$ and $f$, whose transition couples to the cavity mode.  In what follows, we choose $d_{fi}$ as the matrix element connecting the lowest-energy bright transition in the molecular system ({\em i.e.}, from state $|0^+\rangle$ to $|1^-\rangle$, see below and Ref.\cite{fischer2021}). The coupling parameter $g$ (or $\eta$) depends on the geometry of the cavity, on the number of molecules and on the dielectric constant of the material inside the cavity \cite{flick2017a,fischer2021}. Explicitly, we consider only a single molecule here and we assume a non-lossy cavity. We treat $\eta$ (and therefore also $g$ and $\Omega_R$) as a tunable, ``effective'' coupling parameter in this work. For $0<\eta\leq 0.1$, the regime is denoted as vibrational strong coupling (VSC) regime and $\eta>0.1$ defines the vibrational ultrastrong coupling (VUSC) regime according to Ref.\cite{kockum2019}.
\vspace{0.2cm}
\\
The first term of $\hat{H}_\mathrm{SC}$ in Eq.(\ref{hsc}) resembles the bare light-matter interaction and the second contribution, quadratic in the molecular dipole moment
\begin{equation}
\hat{H}_\mathrm{DSE}
=
\dfrac{g^2}{\hbar\omega_c}
{d}^2(q)
\label{eq.dse}
\end{equation}
is the dipole self energy (DSE) term, which is sometimes neglected (see above). For later use, we introduce the notation
\begin{equation}
\Delta\hat{H}_\mathrm{SC}
=
\hat{H}_\mathrm{SC}
-
\hat{H}_\mathrm{DSE}
=
\sqrt{\dfrac{2\omega_c}{\hbar}}\,g\,x_c
{d}(q)
\label{eq.bare_light_matter_interaction}
\end{equation}
for the bare light-matter interaction without the DSE term. Further, the sum of the potential energy contributions in the vibrational Pauli-Fierz Hamiltonian defines the cavity Born-Oppenheimer potential energy surface (cPES),
\begin{equation}
V_\eta(q,x_c)
=
V(q)
+
\dfrac{\omega^2_c}{2}\,x^2_c
+
\sqrt{\dfrac{2\omega_c}{\hbar}}\,g\,x_c\,d(q)
+
\dfrac{g^2}{\hbar\omega_c}\,d^2(q) \quad ,
\label{cpes}
\end{equation}
which depends on $\eta$ \textit{via} Eq.\eqref{eq.g_vs_eta_relation}. The eigenstates of the Pauli-Fierz Hamiltonian, $\chi_n(q,x_c)$, are denoted as vibro-polaritonic states, which have recently been employed by the authors to interpret vibro-polaritonic spectra of the cavity-altered ammonia model problem\cite{fischer2021}. Here instead, we will use this Hamiltonian and the cPES to compute fully quantum mechanical or TST reaction rates. Additionally, we will consider the time-evolution of a vibro-polaritonic wavepacket, $\psi(q,x_c,t)$, governed by the time-dependent Schr\"odinger equation (TDSE)
\begin{equation}
{i}\hbar\dfrac{\partial}{\partial t}
\psi(q,x_c,t)
=
\left(
\hat{H}_\mathrm{S}
+
\hat{H}_\mathrm{C}
+
\hat{H}_\mathrm{SC}
\right)\,
\psi(q,x_c,t)
\label{tdse}
\end{equation}
with initial state $\psi(q,x_c,t_0)=\psi_0(q,x_c)$. In this work, we solve Eq.(\ref{tdse}) {\em via} the multiconfigurational time-dependent Hartree (MCTDH) method\cite{meyer2012} (\textit{cf.} Appendix A).
\subsection{A One-Dimensional Ammonia-Inversion Model}
\label{model}
We study the cavity-altered ammonia inversion using a one-dimensional molecular plus single-cavity mode model, see Ref.\cite{fischer2021}.
\begin{figure}[hbt]
\includegraphics[scale=1.0]{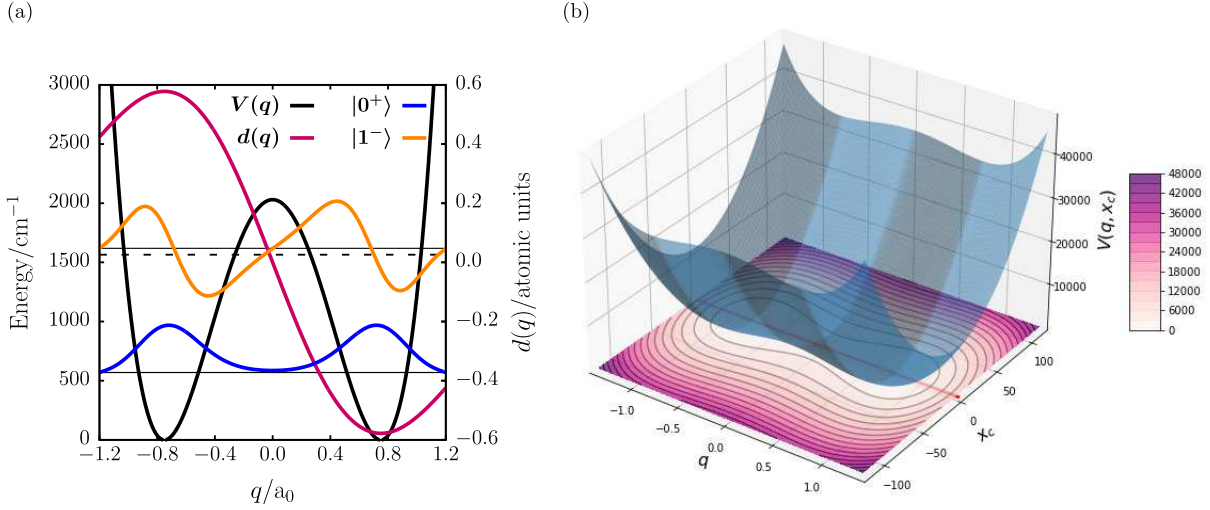}
\renewcommand{\baselinestretch}{1.}
\caption{(a) Symmetric double-well potential, $V(q)$, modeling the ammonia inversion mode, a model dipole function, $d(q)$, as well as the symmetric ground state, $\ket{0^+}$, and the anti-symmetric, first excited state, $\ket{1^-}$ (dashed-black line indicates symmetric, first excited state $\ket{1^+}$). The inversion coordinate and dipole moment are given in atomic units ($a_0$ and $e a_0$ with Bohr radius $a_0$ and elementary charge $e$). (b) Two-dimensional cavity potential energy surface of the non-interacting ($\eta=0$) cavity-plus-ammonia inversion model with minimum energy path (MEP) given in red. Coordinates are given in atomic units ($a_0$ for $q$ and $\sqrt{m_e}\,a_0$ for $x_c$), the potential energy in wavenumbers, $\mathrm{cm}^{-1}$.}
\label{fig.pnh3_cpes_3d_fig}
\end{figure}
The ammonia inversion mode is modeled by a symmetric double-well potential 
\begin{equation}
V(q)
=
A_0
+
A_2\,q^2
+
A_2\,q^4,
\label{v1d}
\end{equation}
as shown in Fig.\ref{fig.pnh3_cpes_3d_fig}(a), with inversion coordinate $q$, measuring the displacement of the H$_3$-plane and the N-atom in ammonia, considered on an interval $q\in[-1.6,1.6]\,a_0$. Further, $A_0=9.249\times10^{-3}\,\mathrm{E}_h=2030\,\mathrm{cm}^{-1}=E^a_{cl}$ is the classical inversion barrier and $A_2=-3.289\times10^{-2}\,{E}_h/a_0^2$ and $A_4=2.923\times10^{-2}\,{E}_h/a_0^4$, respectively. The reduced mass of the ammonia inversion mode is $\mu=m_{3\mathrm{H}}m_\mathrm{N}/m_{\mathrm{NH}_3}=4533.52\,{m}_e$, with electron mass ${m}_e$. Diagonalization of the 1D system Hamiltonian in Eq.\eqref{v1d} gives the lowest tunneling splitting $\hbar\omega_{0^-0^+}=0.92\,\text{cm}^{-1}$, compared to the experimental value of $0.79\,\text{cm}^{-1}$ (superscripts $+$ and $-$ denote even and odd states). 
\vspace{0.2cm}
\\
If not stated otherwise, the cavity mode is chosen resonant to the first bright transition of the molecular system, with $\hbar\omega_c=\hbar\omega_{1^-0^+}=1039\,\text{cm}^{-1}$, which involves the symmetric ground state, $\ket{0^+}$, and the excited antisymmetric state, $\ket{1^-}$, as shown in Fig.\ref{fig.pnh3_cpes_3d_fig}(a). The corresponding transition dipole, $d_{1^-0^+}=0.027\,e\,a_0$, was obtained from the model dipole function (\textit{cf.} Fig.\ref{fig.pnh3_cpes_3d_fig}(a)) 
\begin{equation}
d(q)
=
-\gamma\,q\,e^{-\delta\,q^2},
\label{eq.dipole_ammonia}
\end{equation}
with $\gamma=1.271\,\vert e\vert$ and $\delta=0.8887\,\mathrm{a}^{-2}_0$, respectively.\cite{fischer2021} For vanishing light-matter interaction, $\eta=0$, the two-dimensional cPES, $V_0(q,x_c)=V(q)+\frac{1}{2}\omega^2_c\,x^2_c$ is obtained (\textit{cf.} Fig.\ref{fig.pnh3_cpes_3d_fig}(b)), which obeys $C_{2v}$ symmetry \cite{fischer2021}, and has minima at $q^\pm_0=\pm 0.75\,a_0$ and a transition state at $q^\ddagger=0$. By symmetry, both minima are identical, however, we will refer to $q_0^-=-0.75\,a_0$ as the reactant minimum. We note, at non-zero $\eta$ the symmetry of the cPES is reduced to $C_2$.\cite{fischer2021}
\subsection{Thermal Reaction Rate Theory}
\label{subsec.theory_thermal_rates}
In this work, we calculate thermal reaction rate constants from a time-independent perspective based on both cumulative reaction probabilities (CRP) and Eyring transition state theory (TST), respectively. These scattering-type approaches provide  a sensitive probe of the (cavity) transition state region. Therefore, the corresponding rates should only be interpreted as upper bounds as revival effects due to the bound nature of the double-well potential and also because dissipation effects are ignored in both CRPs and Eyring TST.
\subsubsection{The Cumulative Reaction Probability Approach}
Cumulative reaction probabilities, $N(E,\eta)$, can be used to calculate the thermal rate constant $k(T,\eta)$ of a light-matter hybrid system as\cite{miller1975,miller1983}
\begin{equation}
k(T,\eta)
=
\dfrac{1}{2\pi\hbar\,Q_R(T,\eta)}
\displaystyle\int^\infty_0
\mathrm{d}E\,e^{-\beta E}
N(E,\eta) \quad ,
\label{eq.thermal_crp_rate}
\end{equation}
with inverse temperature $\beta=1/k_BT$ and classical reactant partition function, $Q_R(T,\eta)$. In the present context, a dependence on the light-matter regime indicated by $\eta$ appears in Eq.\eqref{eq.thermal_crp_rate}. For the reactant partition function, we use the (in this case uncritical) harmonic approximation, which for the two bound vibrational reactant modes can be written as, $Q_R(T,\eta)=Q^{(1)}_R(T,\eta)\ Q^{(2)}_R(T,\eta)$, with single-mode harmonic partition functions
\begin{equation}
Q^{(i)}_R(T,\eta)
=
\biggl(
1
-
e^{-\beta\hbar\omega^{(i)}_R}
\biggr)^{-1},\,
i=1,2 \quad .
\label{eq.harmonic_rectant_partition_fct}
\end{equation}
Here, $\omega^{(i)}_R$ is the harmonized reactant frequency of the $i^{\mathrm{th}}$ degree of freedom (DoF). Thermal rate constants as defined in Eq.\eqref{eq.thermal_crp_rate} account through $N(E,\eta)$ for both quantum mechanical tunneling and the anharmonic nature of the reactive cavity transition state region. The cumulative reaction probability is calculated from a discrete-variable-representation absorbing-boundary-conditions (DVR-ABC) approach as\cite{seideman1992}
\begin{equation}
N(E,\eta)
=
\mathrm{tr}
\left\{
\hat{\Gamma}_R\,
\hat{G}(E,\eta)\,
\hat{\Gamma}_P\,
\hat{G}^\dagger(E,\eta)
\right\},
\label{eq.cumulative_reaction_probability}
\end{equation}
where the trace, $\mathrm{tr}\left\{\dots\right\}$, runs over a basis spanning the vibro-polaritonic Hilbert space and $\hat{G}(E,\eta)$ is the Green's function of the vibrational Pauli-Fierz Hamiltonian defined as
\begin{equation}
\hat{G}(E,\eta)
=
\left(
E-\hat{H}+\frac{\text{i}}{2}\,\hat{\Gamma}
\right)^{-1}.
\label{eq.pauli_fierz_greens_fct}
\end{equation}
Here, $\hat{\Gamma}$ is an absorbing potential given by
\begin{equation}
\hat{\Gamma}
=
\hat{\Gamma}_R
+
\hat{\Gamma}_P,
\label{gamtot}
\end{equation}
with reactant, $\hat{\Gamma}_R$, and product absorber, $\hat{\Gamma}_P$, respectively, which are specified in Appendix B.\cite{seideman1992} All operators needed for the CRP are here expressed on a sinc-functions discrete-variable-representation grid.\cite{colbert} Numerical details on the evaluation of Eqs.\eqref{eq.thermal_crp_rate} and \eqref{eq.cumulative_reaction_probability}, including the choice of absorbing boundaries, are provided in {Appendix B}.   
\subsubsection{Eyring Transition State Theory}
We compare the CRP-based thermal rate constants, $k(T,\eta)$, to rates obtained in the well-known approximate framework of harmonic Eyring TST, with \cite{levine2005}
\begin{equation}
k^{\mathrm{TST}}(T,\eta)
=
\dfrac{1}{2\pi\hbar\beta}\,
\dfrac{Q^\ddagger(T,\eta)}{Q_R(T,\eta)}\,
\exp\left[
-\beta\left(E^a_{cl}+E^\ddagger_0(\eta)-E^0_R(\eta)\right)
\right],
\label{eq.thermal_etst_rate}
\end{equation}
where $Q_R(T,\eta)$ is given in harmonic approximation by Eq.(\ref{eq.harmonic_rectant_partition_fct}). Further, $Q^\ddagger(T,\eta)=\left(1-e^{-\beta\hbar\bar{\omega}^\ddagger}\right)^{-1}$ is the classical, harmonic cavity transition state (cTS) vibrational partition function, with cTS bound-state or ``harmonic-valley'' frequency $\bar{\omega}^\ddagger$, which depends on $\eta$ in our model. The frequency $\bar{\omega}^\ddagger$ describes quantized motion perpendicular to the reaction path at the cavity transition state and is obtained as an eigenvalue of the Hessian at the cTS. The other eigenvalue of the Hessian at the cTS in our 2D-model gives an imaginary frequency $\mathrm{i}\,|{\omega}^\ddagger|$ (see below), due its saddle-point character.
\vspace{0.2cm}
\\
The classical activation energy, $E^a_{cl}$, in Eq.(\ref{eq.thermal_etst_rate}) is given by
\begin{equation}
E^a_{cl}
=
V^\ddagger(q,x_c)-V(q_0^-,x_{c0}) \quad ,	
\label{eq.classical_activation_energy}
\end{equation}
where $V^\ddagger(q,x_c)$ is the energy at the cTS and $V(q_0^-,x_{c0})$ the energy at the reactant-well minimum. In the TST rate expression, this energy has to be complemented by the zero-point energy (ZPE) corrections for both the bound harmonic transition state modes, $E^\ddagger_0(\eta)$ and the reactant-well normal modes, $E^0_R(\eta)$, with
\begin{equation}
E^\ddagger_0(\eta)
=
\dfrac{\hbar\bar{\omega}^\ddagger(\eta)}{2}  \quad ,
\hspace{1cm}
E^0_R(\eta)
=
\displaystyle\sum^2_{i=1}\dfrac{\hbar\omega^{(i)}_R(\eta)}{2} \quad .
\label{eq.transition_reactant_zpe}
\end{equation}
It has recently been shown analytically and numerically in Ref.\cite{fischer2021} that, for a length-gauge cPES in the dipole approximation, the classical activation energy, $E^a_{cl}$, is independent of light-matter coupling. This in agreement with independent observations of Ref.\cite{lihuo2021a}, and will again be demonstrated below. As will be shown in Sec.\ref{subsec.cpes_topology}, in the double-harmonic approximation, \textit{i.e.}, harmonically approximated $V(q)$ and linearly approximated $d(q)$, $\omega_R^{(1)}$ and $\omega_R^{(2)}$, therefore also $E_R^0$ and $Q_R$, are also independent of the molecule-cavity coupling strength, $\eta$. Hence, differences in $k^{\mathrm{TST}}(T,\eta)$ for various $\eta$ are dominated by cavity-induced changes in the ZPE of the reactive system at the transition state, $\hbar \bar{\omega}^\ddagger$. Neglecting the ZPE corrections would result in similar rates for all $\eta$, incompatible with experiments.
\vspace{0.2cm}
\\
Unlike quantum mechanical (CRP) rates, $k^{\mathrm{TST}}(T,\eta)$ does not account for tunneling.\cite{anders} To include tunneling approximately, a Wigner correction\cite{tanaka1996}
\begin{equation}
\kappa_W(T,\eta)
=
1
+
\dfrac{1}{24}
\left(
\beta\,
\hbar\vert\omega^\ddagger\vert
\right)^2,
\label{eq.wigner_correction}
\end{equation}
can be introduced, which results in a Wigner-corrected Eyring TST rate
\begin{equation}
k^{\mathrm{TST}}_W(T,\eta)
=
\kappa_W(T,\eta)
\cdot
k^{\mathrm{TST}}(T,\eta) \quad .
\label{eq.wigner_thermal_etst_rate}
\end{equation} 
The Wigner correction $\kappa_W(T,\eta)$ introduces a dependence of the thermal rate constant on the barrier frequency, $\vert\omega^\ddagger\vert$, which is related to the barrier width and depends on the light-matter interaction parameter $\eta$ as will be demonstrated in Sec.\ref{subsec.cpes_topology}.
\section{Results and Discussion}
\label{results}
We analyze the cavity-altered ammonia inversion model from both a time-independent and a time-dependent perspective. In our discussion, we either vary the light-matter interaction strength $\eta$, while keeping the cavity frequency resonant to the fundamental transition of the molecular model, \textit{i.e.}, $\omega_c=\omega_{1^-0^+}$. Alternatively, we vary $\omega_c$ and keeping the light-matter regime fixed \textit{via} $\eta$, respectively. Note that, in the latter scenario, both $g$ and $\omega_c$ have to be adapted consistently according to Eq.\eqref{eq.g_vs_eta_relation} to keep $\eta$ constant, \textit{e.g.}, increasing $\omega_c$ implies an increasing light-matter interaction strength, $g$.
\subsection{Analysis of the Cavity Potential Energy Surface}
\label{subsec.cpes_topology}
\begin{figure}[h!]
\includegraphics[scale=1.0]{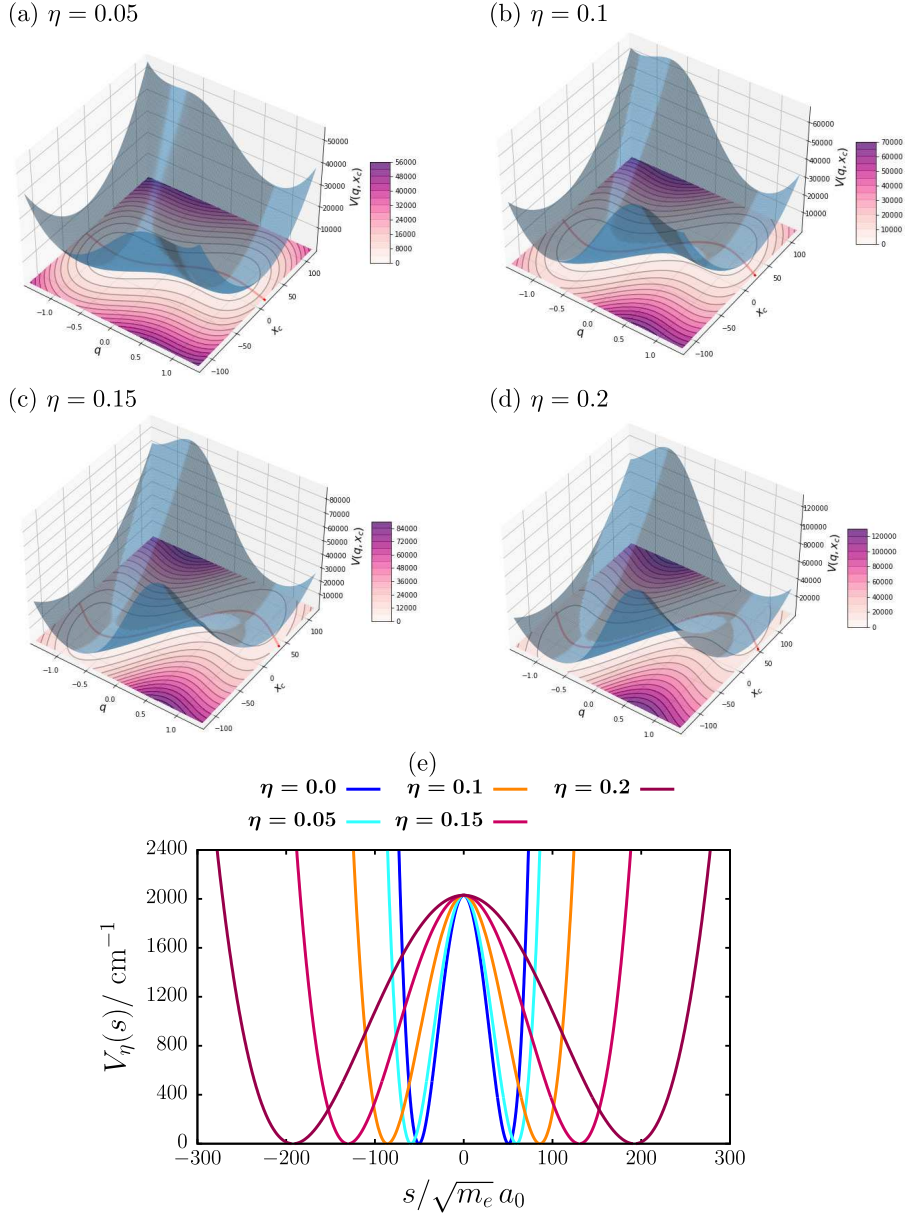}
\renewcommand{\baselinestretch}{1.}
\caption{(a)-(d) Cavity potential energy surfaces for the ammonia inversion model for selected values of $\eta$ at resonant $\omega_c=\omega_{1^-0^+}$, with minimum energy paths indicated by red lines (see text). Coordinates $q$ and $x_c$ are given in atomic units as in Fig.\ref{fig.pnh3_cpes_3d_fig}(a), and the energy scale in wavenumbers, $\mathrm{cm}^{-1}$. (e) Cavity reaction potential, $V_\eta(s)$, along the cavity minimum energy path for selected values of $\eta$ with mass-weighted atomic units for the $s$ coordinate. 
}
\label{fig.pnh3_cpes_interact_fig}
\end{figure}

\begin{figure}[h!]
\includegraphics[scale=1.0]{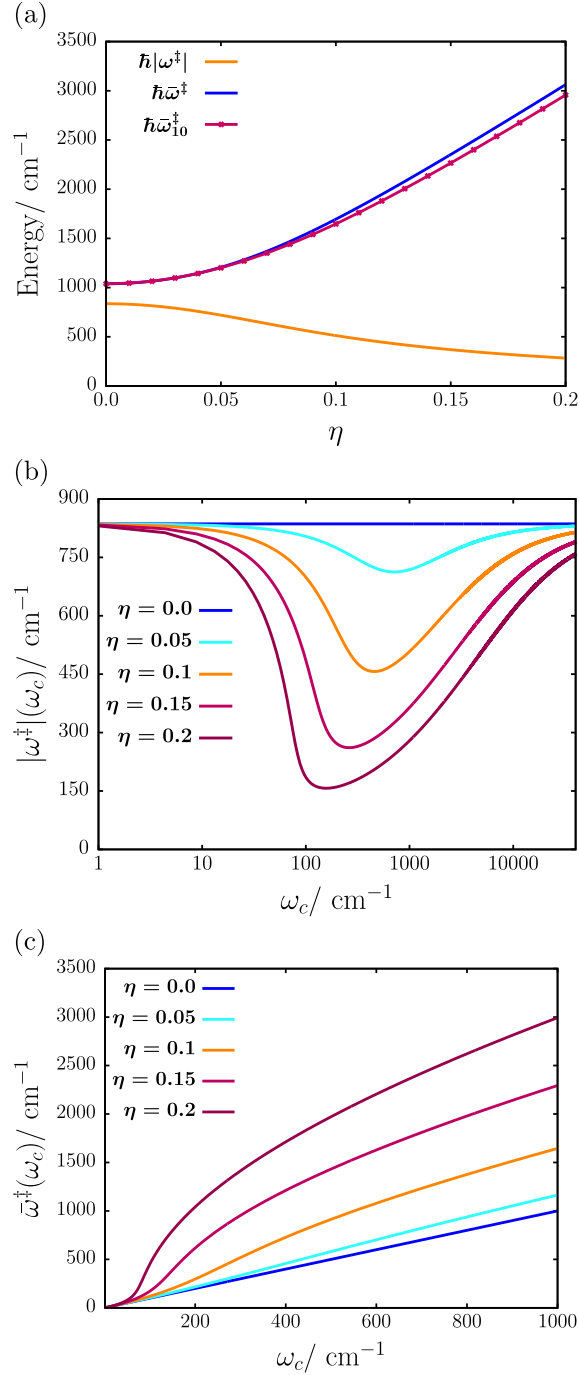}
\renewcommand{\baselinestretch}{1.}
\caption{(a) Harmonic cTS barrier frequency, $\vert \omega^\ddagger\vert$ and 
 ``valley'' frequency $\bar{\omega}^\ddagger$, as well as anharmonic ``valley'' frequency $\bar{\omega}^\ddagger_{10}$ as functions of $\eta$ with $\omega_c=1039\,\mathrm{cm}^{-1}$. (b) and (c): Frequencies $\vert \omega^\ddagger\vert$ (logarithmic $\omega_c$-axis) (b) and  $\bar{\omega}^\ddagger$ (c), as function of cavity mode frequency, $\omega_c$, for selected fixed values of $\eta$.}
\label{fig.pnh3_harmonic_fig}
\end{figure}

We start our discussion by examining the dependence of the cavity PES, $V_\eta(q,x_c)$  ({\em{cf.}} Eq.(\ref{cpes})), on the light-matter interaction strength parameterized by $\eta$. In Fig.\ref{fig.pnh3_cpes_interact_fig}, we show the cPES for selected values of $\eta$.
For increasing $\eta$, we observe a distortion of the cPES, which manifests in the symmetry reduction from the point group $C_{2v}$ in the non-interacting case to the point group $C_2$ at non-zero $\eta$.\cite{fischer2021} Due to symmetry constraints, the molecular coordinates of the double-well minima, $q^\pm_0=\pm0.75\,a_0$ are independent of $\eta$ as is the cTS location at $(q,x_c)=(0,0)$ for all values of $\eta$.
\vspace{0.2cm}
\\
Further, we introduce a cavity minimum energy path (cMEP) as a geodesic curve on the cPES\cite{zhu2019} and study its dependence on the light-matter interaction. Details on its definition and computation for our model system are described in Appendix C. The cMEP is measured by a coordinate $s$, and is indicated by the red lines in Figs.\ref{fig.pnh3_cpes_interact_fig}(a)-(d). 
The coordinate $s$ is employed to evaluate the cavity reaction potential, $V_\eta(s)$, \textit{i.e.}, the cPES $V_\eta(q,x_c)$ evaluated along the cMEP, which is shown in Fig.\ref{fig.pnh3_cpes_interact_fig}(e) for different values of $\eta$. While the double-well structure of the molecular potential is preserved for all light-matter interaction strengths, the potential is stretched along the cavity reaction potential. As a consequence, the barrier along the cMEP {\em broadens} considerably as $\eta$ increases and the barrier frequency $|\omega^\ddagger|$ \textit{decreases} due to the lower curvature at the cTS.
\vspace{0.2cm}
\\
In order to make these observations quantitative, we perform a harmonic analysis of the cTS and the reactant minimum. As shown in Appendix D, in double-harmonic approximation, eigenvalues (and frequencies) can be calculated analytically within our model. For the eigenvalues at the reactant minimum, one finds
\begin{eqnarray}
\omega_R^{(1)} & = & \sqrt{\dfrac{2\,(6A_4\,Q^2_0
+
A_2\,\mu)}
{\mu^2}}, \label{om1} \\
\omega_R^{(2)} & = & \omega_c \quad ,
\label{om2}
\end{eqnarray}
corresponding to two harmonized reactant frequencies. Here, $A_2$ and $A_4$ are the double-well potential parameters of Eq.(\ref{v1d}), and $Q_0= \sqrt{\mu}\,|q_0^-|$ is the mass-weighted coordinate at the reactant minimum. Note that, the frequencies $\omega_R^{(1)}$ and $\omega_R^{(2)}$ are independent of the cavity-molecule interaction $\eta$ as mentioned earlier, because the latter vanishes in second order at the reactant minimum ({\em cf.} Appendix D). With the potential parameters, for the harmonized reactant mode one obtains $\hbar \omega^{(1)}_R=1182\,\mathrm{cm}^{-1}$ from Eq.(\ref{om1}), while $\omega^{(2)}_R$ is the bare, harmonic cavity frequency according to Eq.(\ref{om2}).
\vspace{0.2cm}
\\
Turning to the cavity transition state, the Hessian in double-harmonic approximation reads (\textit{cf.} Appendix D)
\begin{equation}
\underline{\underline{W}}^\ddagger
=
\begin{pmatrix}
\dfrac{2}{\mu_s}
\left(
\dfrac{g^2\,\gamma^2}{\hbar\omega_c}
+
A_2
\right)
&
-\sqrt{\dfrac{2\omega_c}{\hbar\,\mu}}\,g\,\gamma
\vspace{0.2cm}
\\
-
\sqrt{\dfrac{2\omega_c}{\hbar\,\mu}}\,g\,\gamma
&
\omega^2_c\\
\end{pmatrix},
\label{eq.chessian_transition}
\end{equation}
which explicitly depends on the light-matter interaction strength $g$, and on the dipole function parameter $\gamma$ ({\em cf.} Eq.(\ref{eq.dipole_ammonia})). Therefore, also its eigenvalues depend on the cavity-molecule coupling, and so do the resulting cTS frequencies, $\bar{\omega}^\ddagger$ and $\omega^\ddagger$. It can be seen immediately that for vanishing coupling, $g=0$, the frequencies (square roots of the eigenvalues of $\underline{\underline{W}}^\ddagger$) are given by
\begin{eqnarray}
\omega^\ddagger      & = & \mathrm{i}\,  \sqrt{\dfrac{2\,\vert A_2\vert}{\mu}} 
 := \mathrm{i}\, |\omega^\ddagger| \\
\bar{\omega}^\ddagger & = & \omega_c \quad .
\end{eqnarray}
In Fig.\ref{fig.pnh3_harmonic_fig}(a), we show the two cTS hybrid normal mode frequencies, $\omega^\ddagger$ and $\bar{\omega}^\ddagger$, as functions of $\eta$.
The barrier frequency $|\omega^\ddagger|$ of the ``imaginary mode'', substantially decreases from $836\,\mathrm{cm}^{-1}$ at $\eta=0$ to $284\,\mathrm{cm}^{-1}$ at $\eta=0.2$. Since $\omega^\ddagger$ determines the curvature of the cPES along the cavity minimum energy path (cMEP), its decreasing nature leads to an increasing barrier width with increasing $\eta$, in agreement with the $V_\eta(s)$ curves shown in Fig.\ref{fig.pnh3_cpes_interact_fig}(e). Further, the perpendicular, real transition frequency $\bar{\omega}^\ddagger$, increases monotonically from $\omega_c=1039\,\mathrm{cm}^{-1}$ at $\eta=0$ to $3062\,\mathrm{cm}^{-1}$ at $\eta=0.2$. From the cPES perspective, an increase in $\bar{\omega}^\ddagger$ results in a narrowing of the cTS valley. In addition, we show the anharmonic bound state cTS frequency $\bar{\omega}^\ddagger_{10}$ in Fig.\ref{fig.pnh3_harmonic_fig}, which corresponds to the fundamental transition of the anharmonic potential perpendicular to the cMEP at the cTS (\textit{cf.} {Appendix E} for details). There, only minor deviations are observed with respect to $\bar{\omega}^\ddagger$ for all $\eta$ considered, which supports the validity of the harmonic approximation in this model.
\vspace{0.2cm}
\\
We now consider $\omega^\ddagger$ and $\bar{\omega}^\ddagger$ as functions of the cavity frequency $\omega_c$, while keeping $\eta$ constant (\textit{cf.} Figs.\ref{fig.pnh3_harmonic_fig}(b) and (c)). In Fig.\ref{fig.pnh3_harmonic_fig}(b), we observe the formation of a minimum in $\omega^\ddagger$ for $\eta>0$, when the cavity frequency $\omega_c$ is close to the barrier frequency, $\omega_c \sim |\omega^\ddagger|$ (as noted in the introduction, a similar effect has been reported recently in theory by Li \textit{et al.}\cite{lihuo2021a}). As $|\omega^\ddagger|$ is related to the barrier width, this minimum corresponds to a maximum in the barrier width. Such a ``barrier resonance effect'' is, however, not able to explain experiments by Ebbesen and coworkers\cite{thomas2016}, where resonances were found close to reactant normal mode frequencies -- an effect to be further discussed below. 
In contrast to the barrier frequency, the valley frequency, $\bar{\omega}^\ddagger$, increases monotonically with $\omega_c$ for all values of $\eta$ considered here (\textit{cf.} Fig.\ref{fig.pnh3_harmonic_fig}(c)), and no resonance is observed. 
\vspace{0.2cm}
\\
From the cPES analysis, we deduce two consequences relevant for thermal reaction rates, which will be analyzed in the following subsection: First, the ``stiffening'' of the cTS valley with increasing $\eta$ leads to a decrease of energetically accessible states at the cavity transition state, which suggests a decrease of the thermal reaction rates with $\eta$. Second, the broadening of the barrier will suppress tunneling, which further decreases reaction rates in a cavity-altered isomerization reaction. Importantly, both effects are purely quantum mechanical in nature and cannot be observed in a classical setting. 
\subsection{Cavity-Altered Thermal Quantum Isomerization Rates}
\label{subsec.results_thermal_rates}
We address the cPES-topological effects from the perspective of thermal reaction rates and compare results obtained from the cumulative reaction probability approach and harmonic Eyring TST, respectively. We start with harmonic Eyring TST rates $k^{\mathrm{TST}}(T,\eta)$, (\textit{cf.} Eq.(\ref{eq.thermal_etst_rate})), and show  Arrhenius plots $\ln k^{\mathrm{TST}}$ {\em vs.} $1/k_BT=\beta$, for selected values of $\eta$, with cavity frequency $\omega_c=1039$ $\text{cm}^{-1}$. The $\beta$-range shown of [0.001,0.02] cm, corresponds to a temperature range [1438,72] K.
\\

\begin{figure}[hbt]
\includegraphics[scale=1.0]{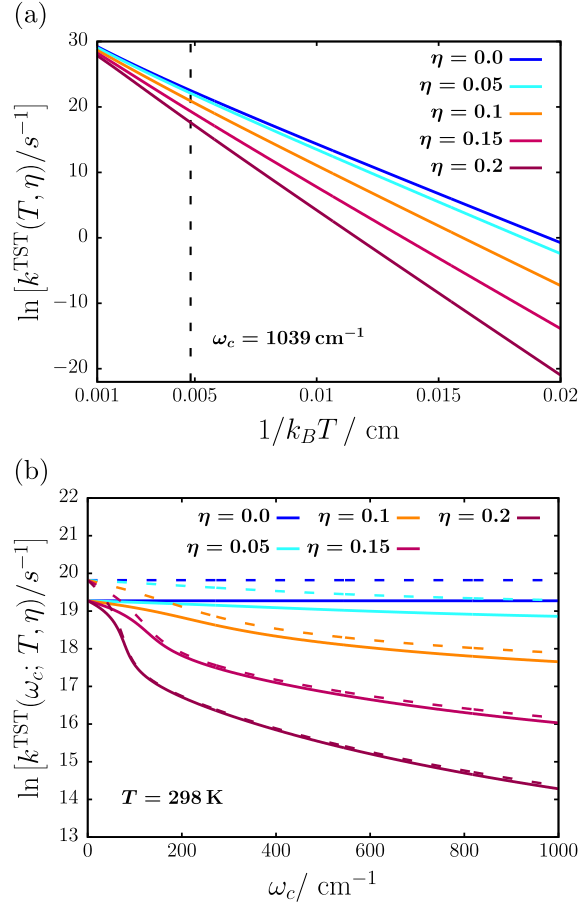}
\renewcommand{\baselinestretch}{1.}
\caption{(a) Natural logarithm of harmonic Eyring TST rates $\ln k^{\mathrm{TST}}(T,\eta)$ as function of inverse temperature $1/k_BT$ at $\omega_c=1039$ $\text{cm}^{-1}$ for selected $\eta$ (black, dashed-line indicates $T=298\,\mathrm{K}$). (b) Natural logarithm of Eyring TST rates $\ln k^{\mathrm{TST}}(\omega_c;\,T,\eta)$ with (dashed) and without (bold) Wigner tunneling correction as a function of the cavity frequency $\omega_c$ for $T=298\,\text{K}$.}
\label{fig.pnh3_rates_resonance_fig}
\end{figure}

\begin{figure}[hbt]
\includegraphics[scale=1.0]{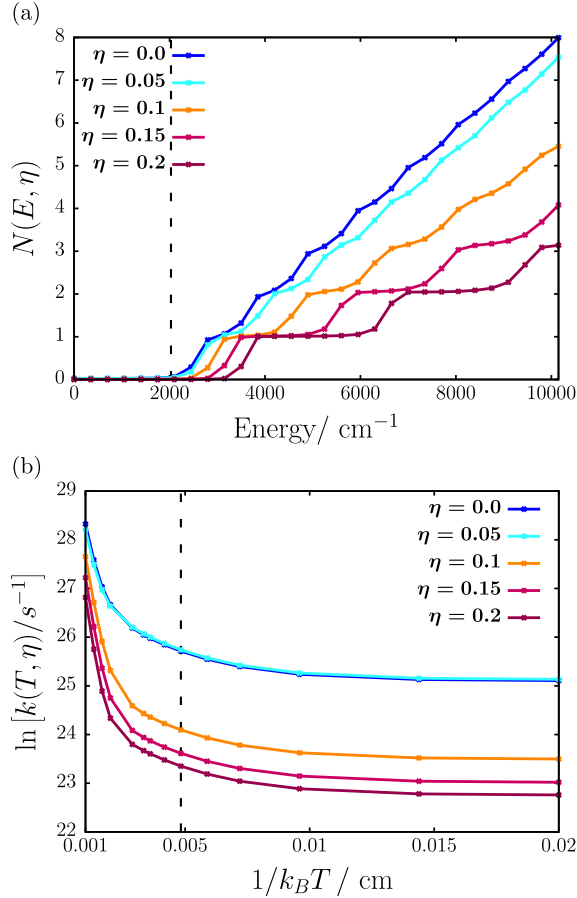}
\renewcommand{\baselinestretch}{1.}
\caption{(a) Cumulative reaction probability $N(E,\eta)$ for the cavity-altered ammonia inversion model as a function of energy $E$ above the reactant well minimum for selected values of $\eta$. The classical barrier height, $E^a_{cl}=2030\,\mathrm{cm}^{-1}$, is indicated by the vertical dashed black line. (b) Arrhenius plot of the  logarithmic rate constants, $\ln k(T,\eta)$, obtained from the CRP, as a function of inverse temperature $1/k_BT$ for selected values of $\eta$ (here black, dashed-line indicates $T=298\,\mathrm{K}$). In both plots (a) and (b), the cavity frequency is $\omega_c=1039$ $\text{cm}^{-1}$.}
\label{fig.pnh3_crp_rates_fig}
\end{figure}

In Fig.\ref{fig.pnh3_rates_resonance_fig}(a), we observe an overall decrease in $k^{\mathrm{TST}}$ with increasing $\eta$, {\em i.e.}, our model agrees qualitatively with an experimentally observed rate-diminishing effect of the cavity.\cite{thomas2019} At high temperatures, the rates converge to a similar value and at low temperatures, we observe suppressed rates for increasing $\eta$, which decrease monotonically with $1/k_BT$. With respect to the topological changes of the cPES (\textit{cf.} Sec.\ref{subsec.cpes_topology}), we deduce that the cavity-induced reduction even of $k^{\mathrm{TST}}$ is of purely quantum mechanical character as it originates from the ZPE corrections given by Eq.\eqref{eq.transition_reactant_zpe}. In particular, the rate reduction is not related to changes in the classical activation barrier energy $E^a_{cl}$, which is independent of $\eta$ as argued in context of the cPES. Despite this independence, a representation of $k^{\mathrm{TST}}(T,\eta)$ in the thermodynamic form of TST as 
\begin{equation}
k^{\mathrm{TST}}(T,\eta) = \dfrac{1}{2\pi\hbar\beta}\, 
 e^{-\beta \Delta G^\ddagger}
 \label{eq.thermodynamics_tst_rates}
\end{equation}
will give activation free energies $\Delta G^\ddagger$, which substantially increase with increasing $\eta$. Denoting the temperature-dependent prefactor in Eq.\eqref{eq.thermal_etst_rate} as $B(T)$, the logarithmic form of Eq.\eqref{eq.thermodynamics_tst_rates} takes the form
\begin{equation}
\ln k^{\mathrm{TST}}(T,\eta) = \ln B(T) - \beta \left(E_a^{cl}-E_R^0+\frac{\hbar \bar{\omega}^\ddagger}{2} \right) \quad .
\end{equation}
Since $B(T)$ depends comparatively weakly on $T$ and on $\eta$, while $E_a^{cl}$ and $E_R^0$ are independent of $\eta$ in our model, the main changes in the slope of $\ln k^{\mathrm{TST}}(T,\eta)$ are due to $\eta$-dependent term $\hbar \bar{\omega}^\ddagger/2$ (\textit{cf.} Fig.\ref{fig.pnh3_harmonic_fig}(a)). Hence, an increase in $\Delta G^\ddagger$ with increasing $\eta$ is solely due to the ZPE of the perpendicular (``valley'') mode at the cTS and not due to an increased classical barrier. In passing, we note that the Arrhenius plots in Fig.\ref{fig.pnh3_rates_resonance_fig}(a) are not strictly linear, due to the remaining $T$-dependence of $B$.
\vspace{0.2cm}
\\
Further, we study the $\omega_c$-dependence of $k^{\mathrm{TST}}(T,\eta)$ at $T=298\,\text{K}$ for fixed values of $\eta$ to investigate the impact of the resonance effect in the (imaginary) barrier frequency $\omega^\ddagger$ (\textit{cf.} Fig.\ref{fig.pnh3_harmonic_fig}(b)). In Fig.\ref{fig.pnh3_rates_resonance_fig}(b), $\ln k^{\mathrm{TST}}(T,\eta)$ curves are shown as a function of $\omega_c$, where bold lines relate to ordinary Eyring TST rates and dashed lines correspond to Wigner tunneling corrected rates (\textit{cf.} Eqs.\eqref{eq.wigner_correction} and \eqref{eq.wigner_thermal_etst_rate}). In both cases, we observe a monotonic decrease with increasing $\omega_c$, with nonlinear rates of change for small $\omega_c$ and a linear regime for large $\omega_c$. The Wigner corrections are comparatively small when couplings $\eta$ and cavity frequencies $\omega_c$ are large, and become slightly more important for smaller $\eta$ in particular when $\omega_c$ is also small.  The only signature of the resonance in $\omega^\ddagger$ as observed in Fig.\ref{fig.pnh3_harmonic_fig}(b), manifests itself in the nonlinear decrease at small values of the cavity frequency, which vanishes with increasing $\omega_c$ where the ZPE corrections start to dominate the behavior of rate constants. Thus, from a time-independent perspective, the barrier resonance effect does not affect quantum mechanically corrected Eyring TST thermal rate constants due to the dominant impact of the ZPE corrections.
\vspace{0.2cm}
\\
To address the effects of tunneling in greater detail while keeping effects of vibrational quantization and anharmonicities fully, quantum mechanical thermal rate constants $k(T,\eta)$ are derived from cumulative reaction probabilities, $N(E,\eta)$, according to Eq.(\ref{eq.thermal_crp_rate}). In Fig.\ref{fig.pnh3_crp_rates_fig}(a), CRPs are shown  as functions of energy $E$ for selected values of $\eta$. The energy axis is given with respect to the reactant well minimum and the classical activation barrier of $E^a_{cl}=2030\,\mathrm{cm}^{-1}$ is indicated by a vertical, dashed black line.
\\

In the non-interacting limit ($\eta=0$), we observe an increase of $N(E,\eta)$ in the vicinity of $E^a_{cl}$ in a step-like fashion, which indicates the successive opening of reactive channels. The number of reactive channels is related to the number of bound states at the cavity transition state.The step structure of $N(E,\eta)$ is found for all values of $\eta$, however, with increasing light-matter interaction strength the number of steps decreases. Additionally, for increasing $\eta$, the slope of $N(E,\eta)$ with energy $E$ decreases significantly over the VSC to the VUSC regime and the onset around $E^a_{cl}$ is blue-shifted. The reduction of reactive channels, \textit{i.e.}, steps in $N(E,\eta)$ is mostly due to a narrowing of the cTS valley. As a consequence, a smaller number of channels is open at a given energy, the larger $\eta$. The distance between two steps is approximately given by $\hbar \bar{\omega}^\ddagger$, reflecting the quantization of the perpendicular mode at the cTS. For instance, for $\eta=0.2$, $\hbar \bar{\omega}^\ddagger \sim 3000$ cm$^{-1}$ according to Fig.\ref{fig.pnh3_harmonic_fig}(a), which is close to the distance of the midpoints of the first and second step of $N(E)$ and for $\eta=0$, $\hbar \bar{\omega}^\ddagger \sim 1000$ cm$^{-1}$ (\textit{cf.} Figs.\ref{fig.pnh3_harmonic_fig}(a) and \ref{fig.pnh3_crp_rates_fig}(a)). We note that, for a given $\eta$, the steps are getting slightly narrower the higher the energy, in particular when $\eta$ is small, which reflects the weak anharmonicity of the ``valley'' potential. 
\vspace{0.2cm}
\\
Turning to the Arrhenius plots in Fig.\ref{fig.pnh3_crp_rates_fig}(b), we find that the individual rates tend to similar values at high temperatures independent of $\eta$, while at low temperature, the CRP rates differ substantially for different light-matter interaction strengths. Here, the tunneling regime manifests as plateaus in $k(T,\eta)$. At intermediate temperatures, we observe an overall reduction of the thermal reaction rate as $\eta$ increases and for very low temperatures, a significantly attenuated tunneling rate is found. While the reduction of the overall rate at intermediate temperature can be interpreted {\em via} the ``valley narrowing'' induced absence of energetically accessible states at the cTS, the rate reduction in the tunneling regime can be related to the barrier broadening (\textit{cf.} Fig.\ref{fig.pnh3_cpes_interact_fig}(e)). In summary, we deduce two key effects from Fig.\ref{fig.pnh3_crp_rates_fig}, which lead to a reduction of reaction rates in cavities and have already been indicated at the end of Sec.\ref{subsec.cpes_topology}: (i) The stiffening of modes perpendicular to the reaction path at the cTS reduces the number of thermally accessible reaction channels, and (ii) the cavity-induced broadening of the barrier suppresses tunneling.
\vspace{0.2cm} 
\\
While we concentrate on ideal models with isolated molecules here, one of many possible effects of an environment and its influence on rates is briefly discussed.
A (dynamical) environment, such as a solvent or other reacting molecules, will lead to fluctuations of the potential parameters entering $V(q)$. To address this point, we consider exemplarily fluctuations of the barrier height, modeled by the parameter $A_0$ in Eq.(\ref{v1d}), while keeping the remaining parameters of the potential, $A_2$ and $A_4$, constant. To account for barrier fluctuations, we use a simple stochastic averaging procedure originally suggested to model environment-induced potential fluctuations for a hydrogen transfer reaction in a protein environment\cite{anders}. Assuming a Gaussian distribution of barrier heights, $A_0$, centered around the standard value $A_0=A_{0,s}=2030$ cm$^{-1}$ with a width $\sigma = 200$ cm$^{-1}$, averaged rates $k_\mathrm{ave}$ are then obtained by weighting individual rates $k(A_0)$ as 
\begin{equation}
k_\mathrm{ave}
=
\int\frac{e^{-(A_0-A_{0,s})^2/2\sigma^2}}{\sigma \sqrt{2 \pi}} \  k(A_0) \ \mathrm{d}A_0
 \quad .
\label{kave}
\end{equation}
As argued in Appendix F, where also further details are given, one obtains Arrhenius plots which are almost indistinguishable from those shown in Fig.\ref{fig.pnh3_crp_rates_fig}(b), for the same coupling strengths and temperature range. Closer inspection shows that only at weak coupling and high temperatures, slightly larger slopes are found for the averaged Arrhenius curves. However, absolute effects are small even then, for instance, $\ln k/\ln k_\mathrm{ave}=1.0018$ for $\eta=0$ and $1/k_B T=0.001$ cm. That does not mean that an environment will have no effect on reaction kinetics in cavities, but it shows a certain robustness of our findings against fluctuations of potential parameters. We finally note that, in particular the inclusion of barrier width fluctuations besides barrier height fluctuations, can provide a more realistic extension of the presented model study.
\vspace{0.2cm} 
\\
Another small but subtle effect of the cavity at small coupling strengths should be mentioned. According to Fig.\ref{fig.pnh3_crp_rates_fig}, at the highest temperatures the rate without coupling ($\eta=0$, dark blue curves) is slightly higher than at low coupling $\eta=0.05$ (light blue), as expected. At very low temperatures, however, the rate in the weakly coupled cavity is slightly larger than without any coupling. This is rewarding, showing that a cavity can also be used to enhance rather than suppress rates. 
\subsection{Cavity-Induced Resonant Dynamical Localization}
\label{subsec.resonant_dynamical_localization}
\subsubsection{A Dynamical Resonance Effect}
\begin{figure}[h!]
\includegraphics[scale=1.0]{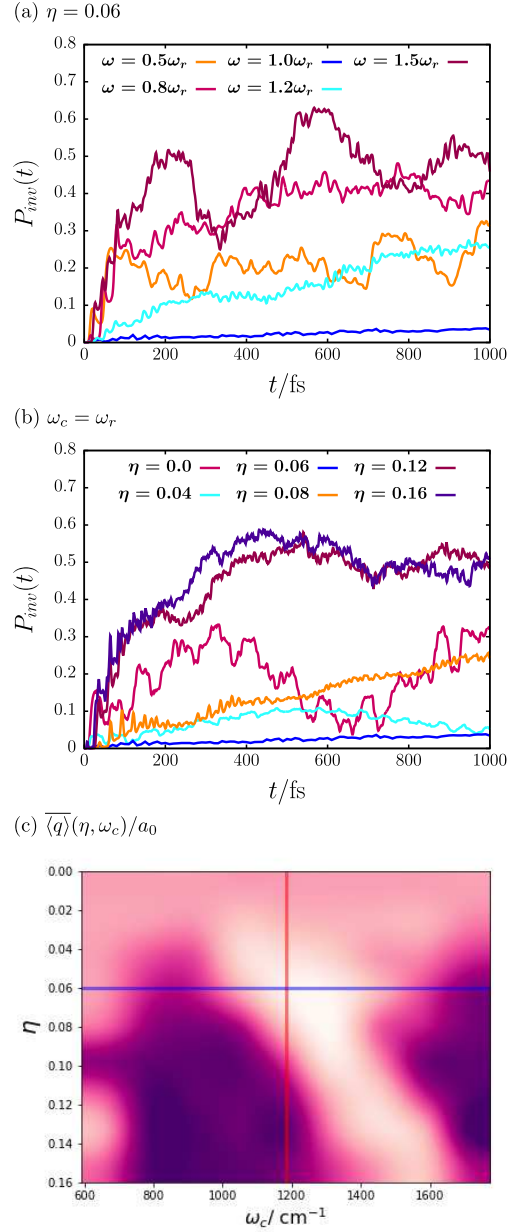}
\renewcommand{\baselinestretch}{1.}
\caption{(a) Inversion probability $P_\mathrm{inv}(t)$ as function of time for selected values of $\omega_c$ at $\eta=0.06$ (blue horizontal 
 line in (c)). (b) Inversion probability $P_\mathrm{inv}(t)$ for selected values of $\eta$ at $\omega_c=\omega_r=1182$ cm$^{-1}$ (red vertical line in (c)). (c) Contour plot of interpolated time-averaged position expectation values $\overline{\braket{{q}}}$ (color bar in atomic units, $a_0$), as function of $\eta$ and $\omega_c$ obtained from $81$ pairs $(\eta,\omega_c)$. 
}
\label{fig.pnh3_resonance_fig}
\end{figure}

In the last part of our discussion, we turn to a time-dependent perspective on the cavity-altered ammonia inversion model and identify a {\em dynamical resonance effect}, leading to a localization effect that reduces inversion probabilities.
For this purpose, we solve the TDSE (\ref{tdse}) via the MCTDH method, starting with an initially ``left-localized'' polariton wavepacket
\begin{equation} 
\psi_0(q,x_c)=\psi_G(q;q_i)\,\phi_0(x_c) \quad .
\end{equation}
Here, $\psi_G(q;q_i)$ is chosen as a molecular Gaussian wavepacket obtained by displacing the ground state of the harmonized reactant mode with frequency $\omega^{(1)}_R=1182\,\mathrm{cm}^{-1}$ to $q_i=-0.9\,a_0$. The latter leads to an initial system energy of $\braket{\hat{H}_\mathrm{S}}=1186\,\mathrm{cm}^{-1}$, which is clearly less than the  classical activation energy, $E^a_{cl}=2030\,\mathrm{cm}^{-1}$. Further, $\phi_0(x_c)$ is the cavity mode ground state, \textit{i.e.}, there are no photons in the cavity mode.
\vspace{0.2cm}
\\
In order to follow the inversion dynamics, we introduce an inversion probability
\begin{equation}
P_\mathrm{inv}(t)
=
\displaystyle\int^\infty_{-\infty}\mathrm{d}x_c
\displaystyle\int^\infty_{-\infty}\mathrm{d}q\,
 \ \theta(q)\,\vert\psi(q,x_c,t)\vert^2 \quad ,
\label{eq.inversion_probability}
\end{equation}
with Heaviside step function $\theta(q)$, which resembles the probability of populating the product (``right'') well specified by $q>0$. In Fig.\ref{fig.pnh3_resonance_fig}(a),(b), we discuss the inversion probability defined in Eq.\eqref{eq.inversion_probability} for two different scenarios: (a) $P_\mathrm{inv}(t)$ for different cavity frequencies, $\omega_c$, given as multiples of the frequency of the harmonized reactant well $\omega_r= \omega_R^{(1)}=1182$ cm$^{-1}$ in the VSC regime with $\eta=0.06$, and (b) $P_\mathrm{inv}(t)$ for various $\eta$, for a fixed cavity frequency $\omega_c=\omega_r$. We follow the dynamics up to  a final time of $t_f=1000\,\mathrm{fs}$ in all cases.
\\

From Fig.\ref{fig.pnh3_resonance_fig}(a), we find that in the ``off-resonant'' scenario, $\omega_c\neq\omega_r$, the inversion probability increases over time, \textit{i.e.}, a significant amount of the vibro-polaritonic wavepacket reaches the product well, with frequently appearing dynamical revivals in the reactant well. In contrast, if the cavity frequency is tuned resonant to the harmonized reactant frequency, $\omega_c=\omega_r$, we observe an overall drastic suppression of $P_\mathrm{inv}(t)$. In this case, we find after $t_f=1000$ fs only $P_{inv}(t_f)=3.5\cdot 10^{-2}$, \textit{i.e.}, less than $4\,\%$ of product forms at resonance. In contrast, one obtains $P_{inv}(t)>0.25$ for other values of $\omega_c$. 
\vspace{0.2cm}
\\
Moreover, from Fig.\ref{fig.pnh3_resonance_fig}(b), we find that at resonance ($\omega_c=\omega_r$), the inversion probability depends quite strongly on the coupling regime indicated by $\eta$. In the VUSC regime  ($\eta>0.1$), the inversion probability is large while in the VSC regime down to the non-interacting limit ($0\leq\eta<0.1$), the inversion yield is lower. However, the detailed behavior of $P_\mathrm{inv}(t_f)$ is quite non-monotonic, with a minimum found for a certain $\eta$ (see below).
\vspace{0.2cm}
\\
The suppressed inversion at some combinations of $\eta$ and $\omega_c$ suggests a localization of the wavepacket in the reactant well around its classical minimum at $q_0^-=-0.75\,a_0$ (\textit{cf.} Fig.\ref{fig.pnh3_cpes_3d_fig}(a)). In order to quantify localization, we consider a time-averaged, molecular displacement expectation value\cite{schaefer2021}
\begin{equation}
\overline{\braket{{q}}}
=
\dfrac{1}{t_f}
\displaystyle\int^{t_f}_0\mathrm{d}t^\prime
\displaystyle\int^\infty_{-\infty}\mathrm{d}x_c
\displaystyle\int^\infty_{-\infty}\mathrm{d}q\ \,
q\,
\vert\psi(q,x_c,t^\prime)\vert^2\quad,
\label{eq.time_averages_position_expect}
\end{equation}  
of the wavepacket $\psi(q,x_c,t)$ as a function of $\omega_c$ and $\eta$. In Fig.\ref{fig.pnh3_resonance_fig}(c), we show $\overline{\braket{{q}}}$ as a contour plot, where the color-bar indicates the magnitude of $\overline{\braket{{q}}}$, which takes values between $-0.7\,a_0\leq \overline{\braket{{q}}}\leq 0.0\,a_0$. The vertical axis resembles the coupling strength, $\eta$,  while the horizontal axis corresponds to the cavity frequency, $\omega_c$. The blue horizontal line relates to $\eta=0.06$ as considered in Fig.\ref{fig.pnh3_resonance_fig}(a) and the red vertical line indicates the harmonized reactant frequency $\omega_r=1182\,\mathrm{cm}^{-1}$ as considered in Fig.\ref{fig.pnh3_resonance_fig}(b).
\vspace{0.2cm}
\\
For small values of $\eta$, the mean displacement shows a rather uniform behavior with respect to variations in the cavity frequency. In the VSC regime for $\eta\geq0.03$, a global minimum forms at $\omega_c$ close to the harmonized reactant frequency $\omega_r$. This resonance exhibits a blue shift as $\eta$ increases. From Fig.\ref{fig.pnh3_resonance_fig}(c), we find the suppressed $P_\mathrm{inv}(t_f)$ at $\eta=0.06$ (\textit{cf.} Fig.\ref{fig.pnh3_resonance_fig}(b)) to be related to a minimum in $\overline{\braket{{q}}}\approx-0.7\,a_0$ close to the reactant minimum at $q_0^-$. Hence, the vibro-polaritonic wavepacket $\psi(q,x_c,t)$ is dynamically localized in the reactant well with respect to the molecular coordinate $q$, which leads to an efficiently suppressed inversion probability $P_\mathrm{inv}(t)$, as shown in Fig.\ref{fig.pnh3_resonance_fig}(a). The small shift of $\overline{\braket{{q}}}$ to larger $q$-values compared to $q_0^-=-0.75\,a_0$ results from the quantum reactant well minimum with respect to the harmonized reactant mode ground state. We observe this ``dynamical'' resonance to be quite sensitive to variations in $\omega_c$ at fixed $\eta$. Both, the observation of a cavity resonance effect in the VSC regime with respect to a reactant normal mode frequency and its sensitivity to changes in the cavity frequency, compare well with a qualitatively identical resonance effect observed by Ebbesen and coworkers\cite{thomas2016}.

\subsubsection{Energetic Perspective on Dynamical Localization}
In order to examine the nature of the molecular mean displacement $\overline{\braket{{q}}}$ in some more detail, we consider the time-evolution of energy expectation values $\langle{\hat{H}_\mathrm{S}}\rangle(t)$, $\langle{\hat{H}_\mathrm{C}}\rangle(t)$, $\langle{\hat{H}_\mathrm{DSE}}\rangle(t)$, $\Delta\langle{\hat{H}_\mathrm{SC}}\rangle(t)$ and $\langle{\hat{H}}\rangle(t)$ in the VSC regime with $\eta=0.06$. This is done for cavity frequencies $\omega_c=0.8\omega_r$ in Fig.\ref{fig.pnh3_resonance_fig}(a), $\omega_c=\omega_r$ in Fig.\ref{fig.pnh3_resonance_fig}(b), and $\omega_c=1.5\omega_r$ in Fig.\ref{fig.pnh3_resonance_fig}(c). 
\\

\begin{figure}[h!]
\includegraphics[scale=1.0]{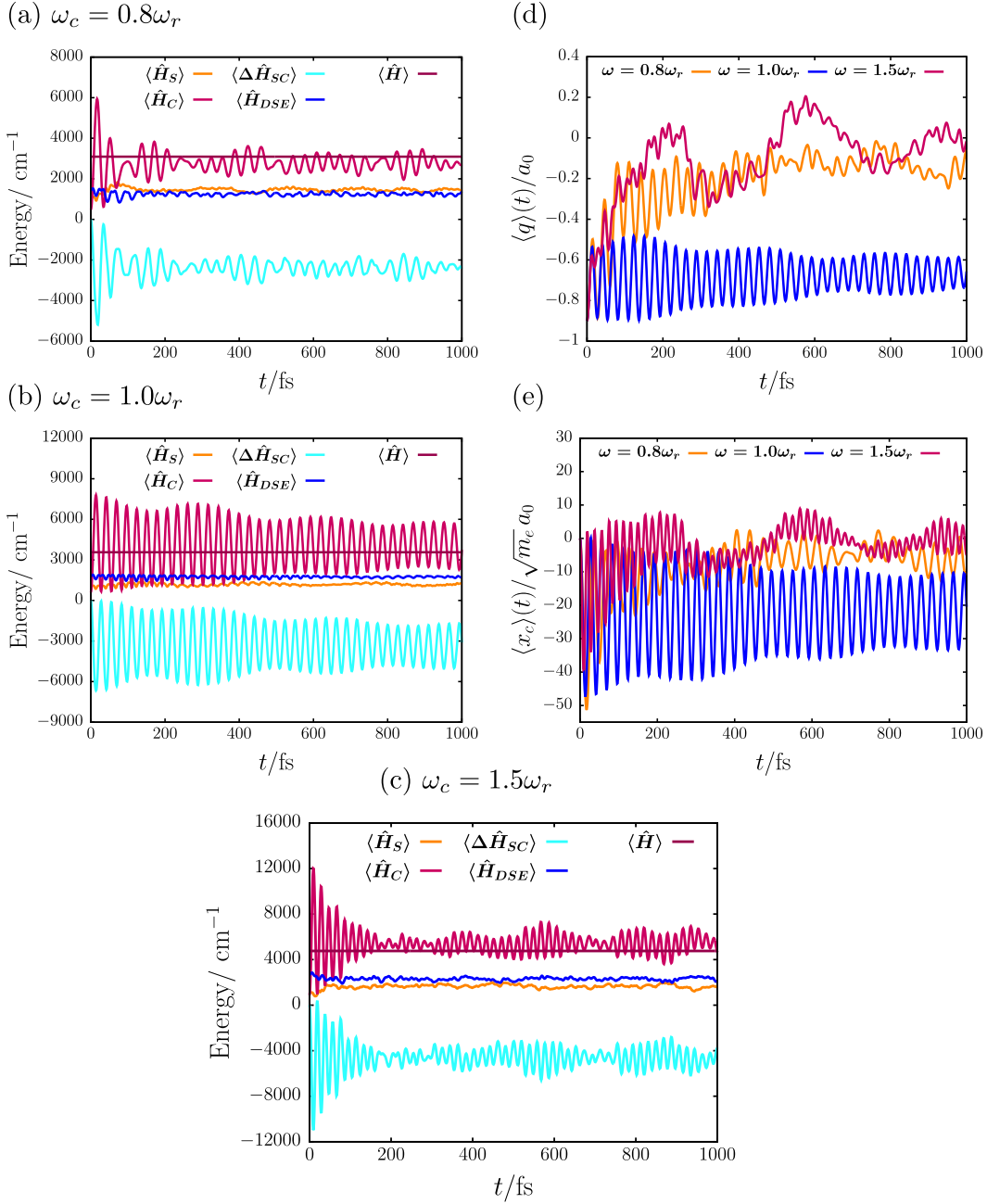}
\renewcommand{\baselinestretch}{1.}
\caption{Time-evolution of energy expectation values $\langle{\hat{H}_\mathrm{S}}\rangle(t)$, $\langle{\hat{H}_\mathrm{C}}\rangle(t)$, $\langle{\hat{H}_\mathrm{DSE}}\rangle(t)$, $\Delta\langle{\hat{H}_\mathrm{SC}}\rangle(t)$ and $\langle{\hat{H}}\rangle(t)$ in the VSC regime with $\eta=0.06$ for cavity frequencies (a) $\omega_c=0.8\omega_r$, (b) $\omega_c=\omega_r$ and (c) $\omega_c=1.5\omega_r$. Expectation values of molecular displacement coordinate $\langle{{q}}\rangle(t)$ (d) and cavity coordinate $\langle{{x}_c}\rangle(t)$ (e) as function of time $t$ for the three cavity frequencies and the VSC regime as considered in (a),(b) and (c). 
}
\label{fig.pnh3_resonance_coord_fig}
\end{figure}

The total energy $\braket{\hat{H}}$ is conserved in our study, as we do not take into account dissipative effects, and is initially given by $\braket{\hat{H}}=\braket{\hat{H}_\mathrm{S}}+\frac{\hbar\omega_c}{2}+\braket{\hat{H}_\mathrm{DSE}}$ (\textit{cf.} Eqs.\eqref{eq.molecular_system} and \eqref{eq.dse}). At times $t>0$, the cavity mode energy $\langle{\hat{H}_\mathrm{C}}\rangle(t)$ and the bare light-matter interaction contribution $\Delta\langle{\hat{H}_\mathrm{SC}}\rangle(t)$ as given by Eq.\eqref{eq.bare_light_matter_interaction} start to oscillate strongly, roughly with opposite phases. In particular, in resonance $\omega_c=\omega_r$, these oscillations show large amplitudes and exhibit a coherent character with small quantum beats. For off-resonant scenarios the dynamics tends in contrast to be incoherent and of lower amplitude. The significant contribution of $\Delta\langle{\hat{H}_\mathrm{SC}}\rangle(t)$, notably in resonance, can be understood in terms of simultaneous excitations of both molecular and cavity mode, which are known as non-rotating wave contributions in the light-matter interaction and become relevant in the VSC regime. Hence, also the system energy $\langle{\hat{H}_\mathrm{S}}\rangle(t)$ oscillates in a non-trivial coherent manner in the resonance scenario, \textit{i.e.}, energy is coherently exchanged between cavity and system, however, at significantly lower amplitude. This coherent effect is fully absent for the off-resonant scenario. Finally, the dipole self-energy $\langle{\hat{H}_\mathrm{DSE}}\rangle(t)$ shows only some small-amplitude dynamics similar to the system. 
\vspace{0.2cm}
\\
Turning to the time-evolution of the molecular, $\langle q \rangle(t)$, and cavity coordinate expectation value, $\langle x_c \rangle(t)$, respectively, we observe a similar strongly oscillatory dynamics in the resonance scenario (\textit{cf.} blue curves in Figs.\ref{fig.pnh3_resonance_fig}(d) and (e)). In particular, the wavepacket exhibits small-amplitude oscillations along the molecular coordinate around the reactant minimum, accompanied by large amplitude motions along the cavity coordinate. Thus, the wavepacket does not effectively reach the transition state region in the resonant case, which would require a significant dynamical component along the molecular coordinate. When comparing to the off-resonant scenario  (\textit{cf.} orange and red curves in Figs.\ref{fig.pnh3_resonance_fig}(d) and (e)), we observe an increase in $\langle q \rangle(t)$ that is accompanied by an increase in $\braket{{x}_c}(t)$, thus indicating a crossing of the cTS located at the origin of the $q$-$x_c$-plane. In combination with the energetic analysis, we attribute the dynamical localization in the reaction coordinate $q$ to a resonantly enhanced energy transfer from the system to the cavity mode. Hence, the wavepacket is strongly excited along the cavity coordinate, \textit{i.e.},  perpendicular to the cavity reaction path, which effectively localizes the system dynamically along the molecular coordinate. 
\vspace{0.2cm}
\\
We finally comment on the relation of our findings to similar results reported recently by Sch\"afer \textit{et al.}\cite{schaefer2021}. As stated in the introduction, these authors predicted a (dynamical) resonance effect for a S$_N$2-reaction, in agreement with experimental work of Ebbesen and coworkers\cite{thomas2016}, based on a multi-dimensional classical-dynamics approach. This effect is qualitatively similar to the one reported in this work. The main difference between our conclusions and those of Ref.\cite{schaefer2021} is that, the latter interpret the rate-diminishing resonance effect as being due to a cavity-mediated collective energy redistribution in the molecular vibrational degrees of freedom (IVR), which effectively lowers the energy in the reactive mode and therefore suppresses the reaction. In contrast, in our fully quantum mechanical but only two-dimensional ammonia-plus-cavity model, the cavity mode acts as a direct acceptor of energy from the system mode leading to a reduction of the inversion probability. Nevertheless, also cavity-mode induced IVR (not included in this work) may contribute in addition. 
\section{Summary and Outlook}
\label{sec.conclusion}
We studied a cavity-altered thermal isomerization model reaction, where we identified a cavity-induced reactive slow-down of the isomerization and a dynamical resonance effect, which manifests itself as a spatial localization of the reactant wavepacket. Our study is based on the vibrational Pauli-Fierz Hamiltonian in length-gauge representation and dipole approximation in the single-cavity mode limit. We investigated a one-dimensional molecular model system for the symmetric inversion mode of ammonia, which was coupled {\em via} a coordinate-dependent dipole function to a single cavity mode in a cavity Born-Oppenheimer framework. The resulting hybrid system was studied for light-matter interactions ranging from the vibrational strong coupling (VSC) to the vibrational ultrastrong coupling (VUSC) regime. Moreover, the cavity mode was considered in coordinate representation, which allowed us to employ the concept of cavity potential energy surfaces (cPESs) to interpret cavity-altered reactivity from a quantum chemical perspective. 
\vspace{0.2cm}
\\
In the first part of our study (\textit{cf.} Sec.\ref{subsec.cpes_topology}), we analyzed the cPES of the two-dimensional model system as a function of the light-matter interaction strength $\eta$ and the cavity mode frequency $\omega_c$. For increasing interaction, we observed a distortion of the cPES, which manifests in a significant broadening of the barrier region along the cavity minimum energy path (cMEP). Further, we found a narrowing of the potential valley perpendicular to the cMEP, at the cavity transition state (cTS), while the classical inversion barrier remained unaffected. Additionally, we observed a barrier resonance effect similar to one recently reported in a comparable  system\cite{lihuo2021a}, showing up as a maximal barrier width at a certain cavity frequency close to the barrier frequency. 
\vspace{0.2cm}
\\
In the second part (\textit{cf.} Sec.\ref{subsec.results_thermal_rates}), the cPES was employed to compute thermal reaction rate constants, obtained from a time-independent, scattering-type  perspective. We considered full-quantum rate theory based on cumulative reaction probabilies (CRP) calculated from a vibrational Pauli-Fierz Green's function, and harmonic Eyring transition state theory (TST), with and without Wigner tunneling corrections. We observed an overall significant reduction of isomerization rates for both approaches due to two quantum effects: (i) A ``stiffening'' of the  perpendicular mode at cTS with increasing light-matter interaction diminishes the number of reactive channels corresponding to thermally accessible states at the cTS, and (ii) a barrier-broadening effect with increasing coupling reduces the tunneling probability in the tunneling regime. Interestingly, in the Wigner tunneling-corrected Eyring approach, no barrier resonance effect was observed in the rates due to the dominance of ``valley-mode'' zero-point energy effects.
\vspace{0.2cm}
\\
In a third part of our work (\textit{cf.} Sec.\ref{subsec.resonant_dynamical_localization}), we investigated the cavity-altered ammonia inversion from a time-dependent perspective and identified a cavity-induced dynamical resonance effect qualitatively similar to results reported experimentally\cite{thomas2016} and recently supported theoretically in Ref.\cite{schaefer2021}. In the VSC regime, the resonance appears at a cavity frequency equal to the harmonized molecular reactant frequency and leads to a dynamical localization of the reactant wavepacket in the reactant well along the molecular coordinate, effectively suppressing the isomerization. 
An energy analysis reveals a strongly coherent energy exchange in the resonance case between system, cavity mode and the photon-matter interaction contribution of the vibrational Pauli-Fierz Hamiltonian, which is absent otherwise. While the vibro-polaritonic hybrid system evolves in time, the cavity mode, initially in its vacuum state, is strongly excited, due to dominant non-rotating wave effects in the light-matter interaction contribution. Hence, the reactant wavepacket otabins a large dynamical component along the cavity coordinate, which results in dynamics perpendicular to the reaction path, thus reducing the inversion probability. 
\vspace{0.2cm}
\\
The present work improves our understanding of the experimentally reported light-induced deceleration of chemical reactions\cite{thomas2019} and resonance effects\cite{thomas2016} in optical cavities. It also offers room for further studies. Among them is an extension to bimolecular reactions for which the scattering-type, time-independent rate models will also be more appropriate. Further, a careful analysis of the effects of few-mode {\em vs.} multi-mode models and quantum {\em vs.} classical models is needed. The different types of resonance phenomena ({\em e.g.}, ``barrier'' {\em vs.} reactant-vibration resonances) should be investigated in more detail. The effect of an environment (\textit{e.g.}, solvent, ensembles of reacting  molecules, lossy cavities) should be included and also the orientation of the molecules w.r.t. to the cavity modes -- of which there can be more than one. Perhaps most importantly, strategies should be developed of how to {\em steer} and {\em increase} the reactivity  in cavities. Steering can be achieved indirectly, through suppressing unwanted side reactions, as demonstrated by Ebbesen and coworkers\cite{thomas2019}. However, more active strategies would be welcome, though, and require most probably also enhanced reactivity. Strategies of cavity-enhanced reactivities (beyond the tiny effect demonstrated in Fig.\ref{fig.pnh3_crp_rates_fig}(b)), could be related to the use of propagating (infrared) photons, which  pump a cavity mode or cavity-modified molecular modes or, more generally, vibro-polaritonic states or wavepackets made thereof, to deliver energy needed to overcome barriers.
\section*{Acknowledgements}
We gratefully acknowledge fruitful discussions with Foudhil Bouakline and Evgenii Titov (both Potsdam) and Inga Ulusoy (Heidelberg). The authors thank the Deutsche Forschungsgemeinschaft (DFG) for financial support through project Sa 547/9.  E.W. Fischer acknowledges support by the International Max Planck Research School for Elementary Processes in Physical Chemistry.
\section*{Data Availability Statement}
The data that support the findings of this study are available from the corresponding author upon reasonable request.
\section*{Appendix A: Numerical Details on MCTDH Calculations}
The MCTDH ansatz\cite{meyer1990,manthe1992,beck2000,meyer2003,meyer2012} for the vibro-polaritonic wavepacket reads
\begin{equation}
\psi(q,x_c,t)
=
\sum^{n_s}_{i_s=1}
\sum^{n_c}_{i_c=1}
A_{i_si_c}(t)
\varphi^{(s)}_{i_s}(q,t)
\varphi^{(c)}_{i_c}(x_c,t),
\end{equation}
with two-dimensional coefficient tensors $A_{i_si_c}(t)$ and orthonormal time-dependent single particle functions (SPFs), $\varphi^{(s)}_{i_s}(q,t)$ and $\varphi^{(c)}_{i_c}(x_c,t)$. The SPFs are subsequently represented in a basis of $m_s,m_c$ primitive basis functions $\{\chi^{(s)}_{l_s}(q),\chi^{(c)}_{l_c}(x_c)\}$, which we chose according to harmonic oscillator discrete variable representation (HO-DVR). The numbers of SPFs, $n_s, n_c$, the number of primitive basis functions, $m_s, m_c$ and the HO-DVR molecular/cavity grid endpoints $q_0/q_f$, $x_{c0},x_{cf}$ are given in Tab.\ref{tab.mctdh_parameters} for different light-matter interaction regimes parametrized by $\eta$. 
\begin{table}[hbt]
    \caption{Number of single particle functions (SPFs) $(n_s,n_c)$, primitive harmonic oscillator DVR basis functions $(m_s,m_c)$ and molecular/cavity grid endpoints $q_0/q_f$, $x_{c0},x_{cf}$ for different light-matter interaction regimes ($\eta$) as consider in MCTDH calculations discussed in Sec.\ref{subsec.resonant_dynamical_localization}.}
        \centering
        \begin{tabular}{|c | c c c c c c c c c c c c|}
                \hline
        $\eta$ && $n_s$ (SPFs) && $n_c$ (SPFs) && $m_s$ (HO-DVR) && $m_c$ (HO-DVR) && $q_0/q_f$ && $x_{c0}/x_{cf}$\\
        \hline\hline
        $0.0-0.08$       && 20  && 20  && 251     && 251 &&  $\left[-1.6,+1.6\right]$ && $\mp316.39$ \\
        $0.1-0.14$        && 24  && 24  && 301     && 301 &&  $\left[-1.6,+1.6\right]$  && $\mp347.61$ \\
        $0.16$  && 26  && 26  && 301     && 301 &&  $\left[-1.6,+1.6\right]$ && $\mp347.61$\\
                \hline
        \end{tabular}
\renewcommand{\baselinestretch}{1.}
    \label{tab.mctdh_parameters}
\end{table}
\\
We employed the MCTDH ansatz as implemented in the \textit{Heidelberg MCTDH package} in its version 8.5.13.\cite{mctdh2019} 

\section*{Appendix B: Details on CRP DVR-ABC Calculations}
To compute cumulative reaction probabilities (CRP), we employ a discrete variable representation (DVR) of the vibrational Pauli-Fierz Hamiltonian $\hat{H}$, the corresponding Green's function $\hat{G}$ and a complex absorbing potential $\hat{\Gamma}$ (CAP). For the two-dimensional cavity-altered ammonia inversion model, the DVR-Hamiltonian matrix elements read
\begin{equation}
\left(\underline{\underline{H}}\right)_{ii^\prime jj^\prime}
=
T^m_{ii^\prime}\delta_{jj^\prime}
+
T^c_{jj^\prime}\delta_{ii^\prime}
+
V_\eta(q_i,x_{cj})\,
\delta_{ii^\prime}\delta_{jj^\prime},
\end{equation}
with $i,i^\prime=1,\dots,N_q$ and $j,j^\prime=1,\dots,N_c$ for the molecular coordinate and cavity coordinate grid, respectively. We employ the Colbert-Miller DVR, with kinetic energy operator given by \cite{colbert}
\begin{equation}
T^a_{ii^\prime}
=
\dfrac{\hbar^2}{2\Delta s^2_a}\,
(-1)^{i-i^\prime}
\begin{cases}
\dfrac{\pi^2}{3},& i=i^\prime
\vspace{0.2em}
\\
\dfrac{2}{(i-i^\prime)^2}, & i\neq i^\prime
\end{cases},
\end{equation}
for coordinates $s_a=(q,x_c)$. Each coordinate is discretized on an equidistant grid with $q_i=i\,\Delta q$, $x_{cj}=j\,\Delta x_c$ and $i,j=0,\pm 1,\pm 2,\dots,\pm\frac{N_s}{2}$ where $N_s=N_q,N_c$ is the number of DVR grid points for the respective DoF. 
\\

The cumulative reaction probability $N(E,\eta)$ is given in a DVR representation as
\begin{equation}
N(E,\eta)
=
\mathrm{tr}\{
\underline{\underline{\Gamma}}_R\,
\underline{\underline{G}}\,
\underline{\underline{\Gamma}}_P\,
\underline{\underline{G}}^\dagger
\}
\end{equation}
with matrix representations of the reactant and product CAPs, $\underline{\underline{\Gamma}}_R$ and $\underline{\underline{\Gamma}}_P$, and the Green's function $\underline{\underline{G}}$, respectively. Here, $\underline{\underline{\Gamma}}_R$ and $\underline{\underline{\Gamma}}_P$ are matrix representations of the reactant and product absorbers $\hat{\Gamma}_R$ and $\hat{\Gamma}_P$, which were introduced in Eq.(\ref{gamtot}). They are
defined as
\begin{equation}
\hat{\Gamma}_P
=
\theta_P[f]\,\hat{\Gamma},
\hspace{0.5cm}
\hat{\Gamma}_R
=
\theta_R[f]\,\hat{\Gamma},
\end{equation}
where $\theta_P[f]=\theta[f(q,x_c)]$ is the Heaviside step function, such that $\theta_R[f]=(1-\theta_P[f])$. The step function depends on a coordinate dependent function $f(q,x_c)$, which specifies a separating surface between reactant and product regions on the cPES {\em via} the condition $f(q,x_c)=0$\cite{seideman1992}. The DVR representation of the absorbing potentials is given by
\begin{align}
\Gamma^R_{ii^\prime jj^\prime}
&=
\theta^R_{ii}\,
\Gamma^m_{ii^\prime}\,
\delta_{jj^\prime}
+
\theta^R_{ii}\,
\Gamma^c_{jj^\prime}
\vspace{0.2cm}
\\
\Gamma^P_{ii^\prime jj^\prime}
&=
\theta^P_{ii}\,
\Gamma^m_{ii^\prime}\,
\delta_{jj^\prime}
+
\theta^P_{ii}\,
\Gamma^c_{jj^\prime}
\end{align}
where we choose $f(q,x_c)=q$ and 
\begin{align}
\Gamma^m(q)
&=
\dfrac{4\,k_0}{1+\exp\left((q_m-q)/k_1\right)}
+
\dfrac{4\,k_0}{1+\exp\left((q_m+q)/k_1\right)} \quad ,
\label{eq.ws_cap}
\vspace{0.4em}
\\
\Gamma^c(x_c)
&=
\Gamma^c_0\,
\dfrac{(x_c-x_{c0})^n}{(x_{cm}-x_{c0})^n} \quad .
\label{eq.pl_cap}
\end{align}
The CAP defines two stripes along the $q$-coordinate, in which it increases from 0 in the interval $[q_0^i,q_0^+]$ in a smoothed-step like fashion to some finite value ($k_0$) at larger $|q|$. (The CAP is also turned on, along $x_c$, in a quartic-power fashion.) After performing the trace, $N(E,\eta)$ can be written as
\begin{equation}
N(E,\eta)
=
\sum^{N}_{i=1}
\sum^{N}_{j=1}
\Gamma^R_{ii}\,
\vert G_{ij}\vert^2\,
\Gamma^P_{jj} \quad ,
\end{equation}
with $N=N_q\,N_c$, \textit{i.e.}, only matrix elements $G_{ij}$ coupling grid points in the reactant and the product strip contribute.
\vspace{0.2cm}
\\
Convergence has been reached with parameters $k_0=0.08\,E_h$, $k_1=0.1\,a_0$, $q_m=0.75\,a_0$ for Eq.\eqref{eq.ws_cap} and $\Gamma^c_0=0.09\,E_h$, $n=4$, $x_{c0}=0.0\ \sqrt{m_e}\,a_0$, $x_{cm}=200\ \sqrt{m_e}\,a_0$ for Eq.\eqref{eq.pl_cap}. The thermal rate constant, $k(T,\eta)$ as defined in Eq.\eqref{eq.thermal_crp_rate}, is numerically evaluated {\em via} the composite trapezoidal rule. An energy interval $[E_0,E_1]$ has been chosen with $E_0=0$ and $E_1=5\,E^a_{cl}=10149\,\mathrm{cm}^{-1}$, and discretized via $N_e=30$ equidistant grid points with spacing $\Delta E=E_1/N_e$. All numerical results for CRP and corresponding rates have been obtained by means of a private code based on the \textsc{NumPy} library, version 1.20, of \textsc{Python} in its recent version 3.8. 
\section*{Appendix C: Definition and computation of the cavity minimum energy path}
We define a cavity minimum energy path (cMEP) as a geodesic curve on the cPES\cite{zhu2019}. The cMEP is given by a vector $\underline{r}_\eta(Q)$ parametrized by the mass-weighted molecular coordinate $Q=\sqrt{\mu}\,q$ as
\begin{equation}
\underline{r}_\eta(Q)
=
\left(
Q,x^{(0)}_c(Q)
\right)^T,
\end{equation}
where $x^{(0)}_c(Q)$ is the cavity coordinate obtained from the minimum condition\cite{fischer2021}
\begin{equation}
\dfrac{\partial}{\partial x_c}V_\eta(Q,x_c)=0
\hspace{0.3cm}
\Longrightarrow
\hspace{0.3cm}
x^{(0)}_c(Q)
=
-\sqrt{\dfrac{2}{\hbar\omega^3_c}}\,g\,d(Q) \quad .
\end{equation}
In Figs.\ref{fig.pnh3_cpes_3d_fig}(b) and \ref{fig.pnh3_cpes_interact_fig}(a)-(d), $\underline{r}_\eta(Q)$ is displayed parametrically by red lines on the cPES. At vanishing or small values of $\eta$, the cMEP is linear and parallel to the q-axis. For increasing $\eta$, it turns into an s-shaped curve, which naturally leads to an ``elongation'' of the path. This elongation of the cMEP can be made quantitative by reparametrization in terms of the arc length $L$ as defined by\cite{zhu2019}
\begin{equation}
L\left[\underline{r}_\eta(Q)\right]
=
\displaystyle\int^{Q_f}_{Q_0}
\sqrt{
\left(
\dfrac{\partial\,r^{(1)}_\eta(Q^\prime)}{\partial Q^\prime}
\right)^2
+
\left(
\dfrac{\partial\,r^{(2)}_\eta(Q^\prime)}{\partial Q^\prime}
\right)^2}\,
\mathrm{d}Q^\prime \quad ,
\end{equation}
where $r^{(i)}_\eta(Q),\,i=1,2$ are the two components of $\underline{r}_\eta(Q)$. The arc length $L$ allows us to introduce a cavity reaction path coordinate $s\in[L_0,L_f]$ with $L_0=L\left[\underline{r}_\eta(Q_0)\right]$ and $L_f=L\left[\underline{r}_\eta(Q_f)\right]$, respectively. The coordinate $s$ is employed to evaluate the cavity reaction potential, $V_\eta(s)$, \textit{i.e.}, the cPES $V_\eta(Q,x_c)$ evaluated along $L$.
\section*{Appendix D: Harmonic Analysis of the Cavity PES}
Here, we give details on the harmonic analysis of the cPES, $V_\eta(q,x_c)$, at the reactant well minimum and the cavity transition state. First, we introduce a mass-weighted reaction coordinate $Q=\sqrt{\mu}\,q$ leading to a symmetric double-well potential of the form
\begin{equation}
V(Q)
=
A_0
+
\dfrac{A_2}{\mu_s}\,Q^2
+
\dfrac{A_4}{\mu^2_s}\,Q^4 \quad .
\label{eq.harmonic_doublewell_massweight}
\end{equation} 
Expanding $V(Q)$ up to second order in $Q$ around the reactant minimum with coordinate $Q_0=\sqrt{\mu_s}\,q_0^-$, leads to
\begin{align}
V^0(Q)
&\approx
V(Q_0)
+
\dfrac{1}{2}\left.\dfrac{\partial^2}{\partial Q^2} V(Q)\right\vert_{Q=Q_0}
(Q-Q_0)^2
\vspace{0.2cm}
\\
&=
\dfrac{1}{2}
\left(
\dfrac{6A_4\,Q^2_0
+
A_2\,\mu_s}
{\mu^2_s}
\right)
(Q-Q_0)^2 \quad ,
\label{eq.harmonic_doublewell_minimum}
\end{align}
with $V(Q_0)=0$, where the harmonic reactant frequency is directly obtained as $\omega^{(1)}_R$ as defined in Eq.(\ref{om1}). Further, in the double-harmonic approximation, the molecular dipole function $d(Q)$ is expanded up to first order around $Q_0$ as
\begin{equation}
d_0(Q_0)
=
-\dfrac{\gamma}{\sqrt{\mu_s}}\,Q_0\,
e^{-\frac{\delta}{\mu_s}\,Q^2_0}
=
const.
\label{eq.harmonic_dipole_min}
\end{equation}
Then, the harmonic approximation of the cPES around the global minimum at $Q_0$ is given by 
\begin{multline} 
V^0_\eta(Q,x_c)
=
\dfrac{1}{2}
\begin{pmatrix}
Q-Q_0 & x_c\\
\end{pmatrix}
\underbrace{
\begin{pmatrix}
\dfrac{2\,(6A_4\,Q^2_0
+
A_2\,\mu_s)}
{\mu^2_s}
&
0\\
0
&
\omega^2_c\\
\end{pmatrix}}_{=\underline{\underline{W}}^0}
\begin{pmatrix}
Q-Q_0\\x_c
\end{pmatrix}
\\
+
\sqrt{\dfrac{2\omega_c}{\hbar}}\,g\,x_c\,
d_0(Q_0)
+
\dfrac{g^2}{\hbar\omega_c}\,
d^2_0(Q_0)
\label{eq.harmonic_matrix_cpes_min}
\end{multline}
with mass-weighted Hessian $\underline{\underline{W}}^0$. Note that in harmonic approximation, $\underline{\underline{W}}^0$ is independent of $\eta$ (or $g$), since there is no light-matter interaction in second order at the reactant minimum. Further, $\underline{\underline{W}}^0$ is diagonal and the square roots of the two eigenvalues (diagonal elements), specifying two orthogonal normal modes, are given by $\omega^{(1)}_R=1182\,\mathrm{cm}^{-1}$ (Eq.(\ref{om1})) and $\omega^{(2)}_R=\omega_c$ (Eq.(\ref{om2})), which correspond to the harmonized reactant mode and the bare harmonic cavity frequency, respectively.
\vspace{0.2cm}
\\
At the transition barrier with coordinate $Q=Q^\ddagger=0$, the harmonic approximation to the molecular symmetric double-well potential reads
\begin{equation}
V^\ddagger(Q))
=
A_0
+
\dfrac{A_2}{\mu_s}\,Q^2 \quad ,
\label{eq.harmonic_doublewell_transition}
\end{equation}
with imaginary harmonic barrier frequency
\begin{equation}
\omega^\ddagger
=
\mathrm{i}\,
\sqrt{\dfrac{2\,\vert A_2\vert}{\mu_s}} 
 = \mathrm{i}\, |\omega^\ddagger| \quad .
\label{eq.harmonic_dwfreq_trans}
\end{equation}
(Note, $A_2$ is negative.) Eq.(\ref{eq.harmonic_dwfreq_trans}) is the imaginary transition state frequency in the absence of cavity-molecule coupling in harmonic approximation. For the molecular dipole moment at the cTS, we find in first order
\begin{equation}
d^\ddagger(Q)
=
-\dfrac{\gamma}{\sqrt{\mu_s}}\,Q \quad .
\label{eq.harmonic_dipole_trans}
\end{equation}
Finally, we obtain as harmonic approximation to the cPES at the caviyty transition state
\begin{equation}
V^\ddagger_\eta(Q,x_c)
=
\dfrac{1}{2}
\begin{pmatrix}
Q & x_c\\
\end{pmatrix}
\underbrace{
\begin{pmatrix}
\dfrac{2}{\mu_s}
\left(
\dfrac{g^2\,\gamma^2}{\hbar\omega_c}
+
A_2
\right)
&
-\sqrt{\dfrac{2\omega_c}{\hbar\,\mu_s}}\,g\,\gamma
\vspace{0.2cm}
\\
-
\sqrt{\dfrac{2\omega_c}{\hbar\,\mu_s}}\,g\,\gamma
&
\omega^2_c\\
\end{pmatrix}}_{=\underline{\underline{W}}^\ddagger}
\begin{pmatrix}
Q\\x_c
\end{pmatrix} \quad ,
\label{eq.harmonic_matrix_cpes_trans}
\end{equation}
where the non-diagonal Hessian $\underline{\underline{W}}^\ddagger$ is a function of the light-matter interaction strength $g$. Diagonalizing $\underline{\underline{W}}^\ddagger$ can trivially be done analytically; taking the square root of the two eigenvalues, gives the imaginary transition state frequency $\omega^\ddagger$ and the ``valley frequency'' $\bar{\omega}^\ddagger$ used in the main text for the general, cavity-molecule coupled case.
\section*{Appendix E: Anharmonic Corrections at Cavity Transition State}
The one-dimensional bound states at the cTS are numerically calculated and the fundamental transition frequency $\bar{\omega}^\ddagger_{10}$ is extracted to examine the anharmonic corrections to $\bar{\omega}^\ddagger$ following Ref.\cite{rietze2017}. We first obtain the orthonormal eigenvectors $\underline{v}^\dagger,\bar{\underline{v}}^\dagger$ of the mass-weighted cPES Hessian $\underline{\underline{W}}^\ddagger$, which parametrically depend on the light-matter interaction strength $g$. A one-dimensional grid for the bound mode is obtained {\em via} displacement along $\bar{\underline{v}}^\dagger$ as
\begin{equation}
\underline{R}_k
=
\underline{R}^\dagger
\pm
k\,
\bar{\underline{v}}^\dagger,
\hspace{0.5cm}
\underline{R}^\dagger
=
(0,0)
\end{equation}
with coordinate tuples $\underline{R}_k=(Q_k,x_{c,k})$ and $k\in[-K_0,+K_0]$. Based on the coordinates $\{\underline{R}_k\}$, we obtain a one-dimensional, discrete cut through the cPES with potential $\bar{V}_\eta(Q_k,x_{c,k})$. We solve a one-dimensional time-independent Schr\"odinger equation
\begin{equation}
\left(
\underline{\underline{T}}_m
+
\bar{\underline{\underline{V}}}_\eta
\right)
\,
\underline{\varphi}_m(\eta)
=
\epsilon_m\,
\underline{\varphi}_m(\eta)
\end{equation}
for the eigenvalues $\epsilon_m$ and corresponding eigenstates $\underline{\varphi}_m(\eta)$, where we employ the Colbert-Miller DVR, and, finally, find $\bar{\omega}^\ddagger_{10}=(\epsilon_1-\epsilon_0)/\hbar$.

\section*{Appendix F: Effects of Gaussian Barrier Distribution}
As mentioned in the main text, one effect of an environment are fluctuating molecular potentials, $V(q)$. Here we consider fluctuating barrier heights $A_0$, their effect simply modeled by a Gaussian distribution function and an incoherent rate averaging procedure according to Eq.(\ref{kave}). The remaining parameters, $A_2$ and $A_4$, are not altered in this discussion. In practice, the integral in Eq.(\ref{kave}) was discretized and  five barrier heights [1830,1930,2030,2130,2230] cm$^{-1}$ were considered, centered around $A_0=2030\,\mathrm{cm}^{-1}$ (width parameter 200 cm$^{-1}$). 
\vspace{0.2cm}
\\
When computing quantum mechanical rates for different coupling strengths $\eta$ and plotting Arrhenius curves, the result (not shown) is hardly distinguishable from Fig.\ref{fig.pnh3_crp_rates_fig}(b). Tab.\ref{tab_lnk} shows, for selected $\eta$ and $1/k_B T$ values, that indeed only at weak coupling and high temperatures (small $1/k_B T$ values), some small differences emerge between averaged and non-averaged cases. This leads to slightly larger slopes of the Arrhenius curves for small $\eta$ and $1/k_B T$, in the averaged case.
\begin{table}[hbt]
    \caption{Natural logarithms of $k_\mathrm{ave}$ and $k$ for selected 
 combinations of coupling strength $\eta$ and inverse temperature $1/k_B T$.}
        \centering
        \begin{tabular}{|ll|cc|}
                \hline
$\eta$ & $1/k_B T$ (cm) &  $\ln k_\mathrm{ave}/s^{-1}$ & $\ln k/s^{-1}$ \\
        \hline\hline
0.0& 0.001 &  28.27& 28.32 \\
0.05& &28.16&28.20\\
0.1&&27.63&27.65\\
0.2&&26.82&26.82\\\hline
0.0&0.002&26.68&26.66\\
0.05 &&26.64&26.64\\
0.1&&25.32&25.32 \\
0.2&&24.35&24.34 \\\hline
0.0&0.005&25.75& 25.72 \\
0.05&&25.76&25.74 \\
0.1&&24.10&24.10 \\
0.2&&23.35&23.35 \\\hline
                \hline
        \end{tabular}
\renewcommand{\baselinestretch}{1.}
    \label{tab_lnk}
\end{table}

\end{document}